\documentclass[oupdraft]{bioupd}
\usepackage{graphicx}
\usepackage{booktabs}
\usepackage{makecell}
\usepackage{threeparttable}
\usepackage{tabulary}
\usepackage{tabularx}
\usepackage{booktabs}
\usepackage{multirow}

\newcommand{\tabitem}{~~\llap{\textbullet}~~}
\graphicspath{{/Users/martinamcmenamin/Documents/PhDwork/PhDLATEX/3/figures/}}

\title{Employing latent variable models to improve efficiency in composite endpoint analysis}

\author{Martina McMenamin$^{\ast1}$, Jessica K. Barrett$^{1}$, Anna Berglind$^{2}$, James M.S. Wason$^{1,3}$\\
\textit{[1] MRC Biostatistics Unit, School of Clinical Medicine,
Cambridge Institute of Public Health
Forvie Site, Robinson Way,
Cambridge Biomedical Campus
Cambridge, UK}
\\
\textit{[2] Global Medicines Development, Biometrics and Information Sciences, AstraZeneca, Gothenburg, Sweden}\\
\textit{[3] Institute of Health and Society, Newcastle University, Newcastle, UK}\\
}

\renewcommand\rightmark{Latent variable modelling for composite endpoint analysis}
\renewcommand\leftmark{McMenamin and others}
\begin{document}
\maketitle

\footnotetext{To whom correspondence should be addressed martina.mcmenamin@mrc-bsu.cam.ac.uk}

\begin{abstract}
{Composite endpoints that combine multiple outcomes on different scales are common in clinical trials, particularly in chronic conditions. In many of these cases, patients will have to cross a predefined responder threshold in each of the outcomes to be classed as a responder overall. One instance of this occurs in systemic lupus erythematosus (SLE), where the responder endpoint combines two continuous, one ordinal and one binary measure. The overall binary responder endpoint is typically analysed using logistic regression, resulting in a substantial loss of information. We propose a latent variable model for the SLE endpoint, which assumes that the discrete outcomes are manifestations of latent continuous measures and can proceed to jointly model the components of the composite. We perform a simulation study and find the method to offer large efficiency gains over the standard analysis. We find that the magnitude of the precision gains are highly dependent on which components are driving response. Bias is introduced when joint normality assumptions are not satisfied, which we correct for using a bootstrap procedure. The method is applied to the Phase IIb MUSE trial in patients with moderate to severe SLE. We show that it estimates the treatment effect 2.5 times more precisely, offering a 60\% reduction in required sample size.}
{Latent variable models; Composite endpoints; Responder analysis; Systemic lupus erythematosus}
\end{abstract}

\section{Introduction}
\label{sec1}
Composite endpoints combine multiple outcomes in order to determine the effectiveness or efficacy of a treatment for a given disease. They are typically recommended when a disease is complex or multi-system and meaningful improvement cannot be captured in a single outcome. Furthermore, the endpoint may be a combination of continuous and discrete outcomes which are collapsed in to a single binary responder index.\\ 
Table \ref{compositeeg} shows examples of diseases that use composite endpoints combining multiple continuous and discrete components. Responders in fibromyalgia must respond in two continuous and one ordinal component however responders in trials for frailty or soft tissue infections must respond in a total of five continuous and discrete components. Generally, these composite responder endpoints will be treated as a single binary outcome and analysed using a logistic regression model, which we term the standard binary method. This solves problems with multiplicity however results in large losses in efficiency (\cite{WasonSeaman}). The aim of this paper is to propose a joint modelling framework within which we can model the components of the composite, retaining the information on the original scales of the outcomes, hence increasing efficiency. 
\begin{table}[h!]
\caption{Examples of diseases that use complex composite endpoints combining multiple discrete and continuous measures to determine effectiveness of a treatment including criteria for response and how each component is measured}\label{compositeeg}
\centering
\begin{tabular}{lll}
\hline
\rule[-1ex]{0pt}{4ex}  Disease & Responder endpoint & Measured by   \\
\hline
& & \\
Fibromyalgia & \tabitem achieved a 30\% improvement in & Electronic diary\\
& pain & \\
& \tabitem 30\% improvement in functional & Subscale of Fibromyalgia Impact \\
&status & Questionnaire (FIQ)\\
 &\tabitem improved, much improved, or & 7-point Patient Global Impression of \\
 & very much improved & Change (PGIC) scale\\
  &  &\\
 Frailty &\tabitem BMI$<$18.5 kg/m2 OR $>$10\% & weight and height\\
 & weight loss since last wave &\\
 & \tabitem One positive answer to exhaustion & CES-D questionnaire  \\
 & questions  & \\
 & \tabitem Low grip strength (M $<$ 31.12 kg, & Eg. Jamar hand dynamometer\\
 & F $<$ 17.60 kg) & \\

 & \tabitem Gait speed (M $<$ 0.691 m/s,& Distance/time \\
 & F $<$ 0.619 m/s)  & \\
& \tabitem Low activity (M $<$ 16.5 activity & Activity units derived using intensity \\
& units, F $<$ 13.5 activity units) & vs. frequency \\
&&\\
Necrotizing & \tabitem Alive until day 28 & yes/no \\
Soft Tissue & \tabitem Day 14 debridements $\leq$ 3 & surface area\\
 Infections & \tabitem No amputation if debridement & yes/no \\
& \tabitem  Day 14 mSOFA score $\leq$ 1 & mSOFA score - composite additively \\
& & combining scores in different systems\\
& \tabitem Reduction of at least 3 score& mSOFA score - composite additively \\
& points in mSOFA score& combining scores in different systems\\
& &\\
Systemic lupus  & \tabitem Change in SLEDAI $\leq$ -4 & SLE Disease Activity Index\\
 erythematosus &\tabitem Change in PGA $<$ 0.3 & Physicians Global Assessment\\
& \tabitem No Grade A or more than one & British Isles Lupus Assessment Group \\
 & Grade B in BILAG & \\
 & \tabitem Reduction in oral corticosteroids& Notes \\
 &  &\\
\hline\\
\end{tabular}
\end{table}
One likelihood based method for handling mixed data is the factorisation model. The objective is to factorise the joint distribution and fit a univariate model to each component of the factorisation (\cite{Mixeddata}). This accounts for correlations between the outcomes by including one response as a covariate in the model for the other response. In the graphical modelling literature this has been termed the `Conditional Gaussian Distribution' (\cite{Whittaker, Lauritzen}). An advantage of these methods in relation to the composite endpoint problem is that we may account for correlations between measurements whilst making inference directly on the outcomes that we have measured, hence they fall within a broader class of \lq{direct methods\rq}. Examples of applications of these ideas, which build on the work of \cite{Olkin}, include developmental toxicity studies by \cite{FitzLaird} and in the longitudinal setting, the augmented binary method was developed for application to composite endpoints in clinical trials where the composite is formed of one continuous and one binary outcome (\cite{WasonSeaman, WasonJenkins, McMenamin}). One difficulty with these methods beyond the bivariate scenario is the range of possibilities for the factorisations, with no consensus on how this should be determined. In the case of the SLE responder endpoint with four components, this amounts to 24 possible factorisations, each of which may result in different conclusions (\cite{multreview, Mixeddata}).\\ 
Another family of models used to model mixed outcome types which feature frequently in economics and finance are copulas. These are functions that join or couple multivariate distribution functions to their uniform one-dimensional marginal distribution functions as discussed by \cite{Nelsen1999}. Copulas offer a flexible framework in this setting, as the marginal distribution functions need not come from the same parametric family. While the construction of copulas is considered to be mathematically elegant and the flexibility with which we can model appealing, they are not without their shortcomings. Extensions beyond the bivariate setting are difficult and have failed to perform well in many applications (\cite{Mixeddata}). Other practical implications include poor out-of-sample predictions due to the wide variety of copulas available. These restrictions, along with difficulties in longitudinal settings with unbalanced data structures, have seen few applications of copulas for mixed outcome types in the medical statistics literature (\cite{multreview}). Applications of copulas in mixed outcome settings include \cite{latentcopula} and \cite{copula}.\\
Another likelihood based method that allows for more flexibility when modelling the correlations between outcomes falls within the framework of latent variable models (\cite{GLLAMM}). The multiple outcomes are assumed to be physical manifestations of some underlying latent process. This is modelled by including the same latent variable in each of the models for the observed responses. The outcomes are then assumed to be independent conditional on this latent variable. This solves the problem of deciding the order of factorisations in previously discussed methods however this formulation results in the inclusion of some covariance parameters in the mean structure, leaving the model sensitive to misspecification of the correlation structure (\cite{Sammel2}). One example of these models is introduced by \cite{Sammel}, where effects of covariates of interest are modelled through this shared latent variable. Although these models have the intuitive interpretation that each outcome is attempting to capture underlying disease activity, the correlation matrix is restricted to allow for the same correlation between each pair of outcomes, which is unlikely in practice. This structure is relaxed in work by \cite{Dunson}, where the effects of covariates are included in the model separately from the latent variable. The correlation structure can be further relaxed to allow for a different latent variable for each outcome, meaning that pairs of outcomes are not assumed to have the same correlation. However these models would require integrating out each of the latent variables in order to obtain the joint distribution of interest (\cite{McCulloch}). Furthermore, they are relevant in applications with multiple time points however less so for a single time point, as is the case for the composite endpoint problem. \\
Latent variables have also been used in the setting of mixed continuous and discrete variables to a different end. Namely, the outcomes adopt a correlated Gaussian distribution by assuming that the discrete outcomes are coarsely measured manifestations of underlying continuous variables subject to some threshold specifications, as seen in \cite{multprobit} and \cite{multprobit2}. Specifying discrete variables in terms of a partitioning of the latent variable space into non-overlapping intervals dates back to \cite{Pearson} in relation to his generalised theory of alternative inheritance and has received much consideration in the literature since. Terminology surrounding these models is inconsistent but they are often referred to as multivariate probit models (\cite{multprobit}). This theory can also be found to underpin conditional grouped continuous models (\cite{PoonLee}). By formulating the distribution in this way, we can correlate the error terms between models and work within the familiar paradigm of Gaussian distributions and maximum likelihood theory. The theory and application of these ideas for a mixture of continuous and binary outcomes has featured in the statistics literature, see for example work by \cite{Tate}, \cite{Cox} and \cite{CatalanoRyan}. Generalisations of these ideas, which appear less frequently in the literature, lead to methods for modelling continuous and ordinal variables, with applications in developmental toxicology (\cite{ordinal1, ordinal3, ordinal4, Regan, Faes}). Despite the advantages, the multivariate probit model has not realised its full potential in the applied biostatistics literature. This was noted by \cite{multprobit3} and we believe it still to be the case today. The few applications that do appear tend to demonstrate bivariate scenarios or those that mix continuous and binary or continuous and ordinal, rather than all three. Furthermore, these models have not been considered specifically to address challenges with modelling composite endpoints. Other work has combined thresholding the response variables and introducing latent variables in the model, examples include work by \cite{probit1}, \cite{Mixeddata}, \cite{ordinallatent} and \cite{ordinal2}, however these ideas are most applicable in the longitudinal setting. We will therefore employ the latent outcome framework and investigate its use for the composite endpoint problem. \\
The paper proceeds as follows. In Section 2 we discuss systemic lupus erythematosus (SLE), the motivating example for the methods. In Section 3 we introduce the latent variable model and discuss how we conduct estimation and inference. In Section 4 we compare the behaviour of the latent variable model with the augmented binary and standard binary methods, including the case when the key assumptions are not satisfied. In Section 5 we apply the methods to the Phase IIb MUSE trial in patients with moderate to severe SLE. Finally, in Section 6 we discuss our findings and make some recommendations. 

\section{Motivating example}
\begin{figure}[h!]
\centering \includegraphics[scale=1]{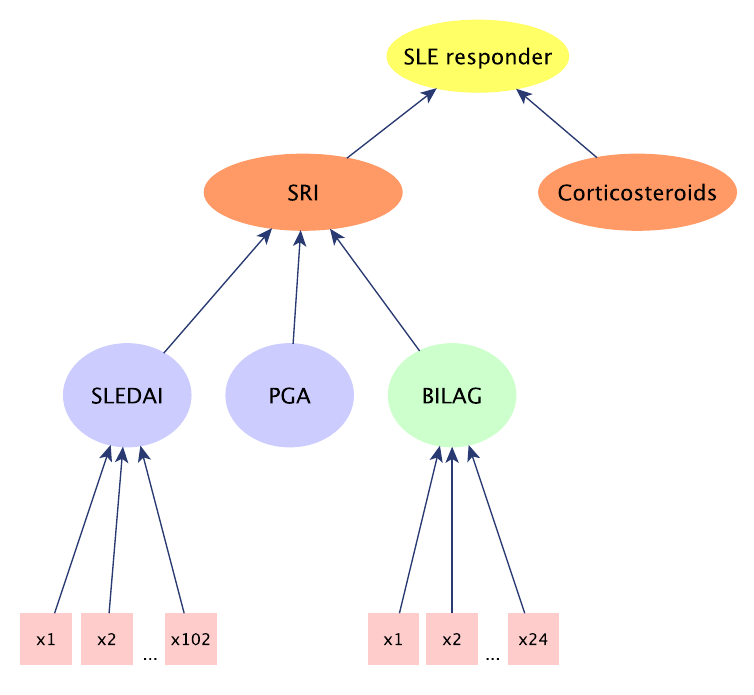}\caption{Structure of the composite endpoint use in trials of systemic lupus erythematosus. The continuous SLEDAI, continuous PGA and ordinal BILAG measures are dichotomised and combined to form the binary SRI indicator which is then combined with the binary taper variable to form the overall binary SLE responder index}\label{SLEendpoint}
\end{figure}

In what follows we will focus specifically on systemic lupus erythematosus (SLE), however the methods introduced will be relevant to other diseases using endpoints with a similar structure. The SLE endpoint is shown in Figure \ref{SLEendpoint}. It combines a continuous PGA measure, a continuous SLEDAI measure, an ordinal BILAG measure and a binary corticosteroids measure, where patients must meet the response criteria in all components in order to be classed as a responder overall. Note that the SLEDAI and BILAG measures are themselves composite scores deriving from a combination of items, however this will not be considered in the analysis. \\
The real data underpinning this motivation comes from the MUSE study (\cite{MUSE}). It was a Phase IIb, randomised, double-blind, placebo-controlled study investigating the efficacy and safety of anifrolumab in adults with moderate to severe SLE. Patients (n=305) were randomised to receive anifrolumab (300mg or 1000mg) or placebo, in addition to standard therapy every 4 weeks for 48 weeks. The primary end point was the percentage of patients achieving an SRI response at week 24 with sustained reduction of oral corticosteroids ($<$10mg/day and less than or equal to the dose at week 1 from week 12 through 24). The methods discussed will make inference at one time point, as this is the case in the trial, although they can be easily extended for the longitudinal case.
\section{Methods}
\subsection{Notation}
Let $\mathbf{Y}_{i}=(Y_{i1}^{}, Y_{i2}^{}, Y_{i3}^{*}, Y_{i4}^{*})$ represent the vector of observed and latent continuous measures for patient i. $Y_{i1}$ and $Y_{i2}$ are the observed continuous SLEDAI and PGA measures. Let $Y_{i3}$ denote BILAG, the observed ordinal manifestation of $Y_{i3}^{*}$ and $Y_{i4}$ the observed binary taper variable for $Y_{i4}^{*}$. $T_{i}$ represents the treatment indicator for patient i, $y_{i10}$ and $y_{i20}$ are the baseline measures for $Y_{i1}$ and $Y_{i2}$ respectively. 

\subsection{Model}
The mean structure for the outcomes is shown in (\ref{model1}). The baseline measures $y_{10}$ and $y_{20}$ are included in the model for $Y_{1}$ and $Y_{2}$ respectively. 

\begin{flalign}\label{model1}
\begin{split}
Y_{i1}&=\alpha_{0}+\alpha_{1}T_{i}+\alpha_{2}y_{i10}+\varepsilon_{i1}\\
Y_{i2}&=\beta_{0}+\beta_{1}T_{i}+\beta_{2}y_{i20}+\varepsilon_{i2}\\
Y_{i3}^{*}&=\gamma_{1}T_{i}+\varepsilon_{i3}^{*}\\
Y_{i4}^{*}&=\psi_{0}+\psi_{1}T_{i}+\varepsilon_{i4}^{*}
\end{split}
\end{flalign}

The observed discrete variables are related to the latent continuous variables by partitioning the latent variable space, as shown in (\ref{discrete}). The lower and upper thresholds for both discrete variables are set at $\tau_{03}=\tau_{04}=-\infty, \tau_{53}=\tau_{24}=\infty$. The intercept term for the ordinal variable in (\ref{model1}) is set at $\gamma_{0}=0$ so that the cut-points $\tau_{13},\tau_{23},\tau_{33},\tau_{43}$ may be estimated. The intercept for the binary outcome $\psi_{0}$ may be estimated, as $\tau_{14}=0$.\\

\begin{align}\label{discrete}
Y_{i3} = \begin{cases}$
Grade E$ & \text{if }  \tau_{03}\leq Y_{i3}^{*}<\tau_{13}, \\
 $ Grade D $& \text{if } \tau_{13}\leq Y_{i3}^{*}<\tau_{23},\\
 $ Grade C$& \text{if } \tau_{23}\leq Y_{i3}^{*}<\tau_{33}, \\
 $ Grade B $& \text{if } \tau_{33}\leq Y_{i3}^{*}<\tau_{43}, \\
 $ Non-responder$& \text{if } \tau_{43}\leq Y_{i3}^{*}<\tau_{53}
\end{cases}&&
Y_{i4} = \begin{cases}
  0, & \text{if } \tau_{04}\leq  Y_{i4}^{*}< \tau_{14}, \\
  1 , & \text{if } \tau_{14}\leq Y_{i4}^{*}< \tau_{24}
\end{cases}
\end{align}\\

Following these assumptions, we can model the error terms in (\ref{model1}) as multivariate normal with zero mean and variance-covariance matrix $\Sigma$, as shown in (\ref{variance}). Note that the error variances for $\varepsilon_{3}^{*}, \varepsilon_{4}^{*}$ are $\sigma_{3}=1$ and $\sigma_{4}=1$. This does not represent a constraint on the model but rather a rescaling required for identifiability. 

\begin{align}\label{variance}
(\varepsilon_{i1}^{}, \varepsilon_{i2}^{}, \varepsilon_{i3}^{*}, \varepsilon_{i4}^{*}) \sim N(0, \Sigma) &&
\Sigma=\begin{pmatrix}
\sigma_{1}^{2} & \rho_{12}\sigma_{1}\sigma_{2} & \rho_{13}\sigma_{1} & \rho_{14}\sigma_{1} \\
\rho_{12}\sigma_{1}\sigma_{2} & \sigma_{2}^{2} & \rho_{23}\sigma_{2} & \rho_{24}\sigma_{2} \\
\rho_{13}\sigma_{1} &  \rho_{23}\sigma_{2} & 1 & \rho_{34} \\
\rho_{14}\sigma_{1} & \rho_{24}\sigma_{2} & \rho_{34} & 1
\end{pmatrix}
\end{align}

Subsequently, we may factorise the joint likelihood contribution for patient i as shown below.

\begin{equation}
l(\boldsymbol{\theta};\mathbf{Y})=f(Y_{i1}^{},Y_{i2}^{};\boldsymbol{\theta})f(Y_{i3}^{*},Y_{i4}^{*}|Y_{i1}^{},Y_{i2}^{};\boldsymbol{\theta} )
\end{equation}

where $\boldsymbol{\theta}$ is a vector which contains all model parameters. The observed likelihood can then be expressed as in (\ref{obslike}). 

%$\boldsymbol{\theta}=(\alpha_{0},\alpha_{1},\alpha_{2},\beta_{0},\beta_{1},\beta_{2},\gamma_{1},\psi_{0},\psi_{1},\sigma_{1},\sigma_{2},\rho_{12},\rho_{13},\rho_{14},\rho_{23},\rho_{24},\rho_{34},\tau_{13},\tau_{23},\tau_{33},\tau_{43})$\\
%l(\theta;Y_{i1},Y_{i2},Y_{i3},Y_{i4})= \\ \frac{1}{2\pi\sigma_{1}\sigma_{2}\sqrt{1-\rho_{12}^{2}}}exp \left\lbrace - \frac{1}{2(1-\rho_{12}^{2})} \left(
%\left(\frac{Y_{i1}-\mu_{1}}{\sigma_{1}}\right)^{2}-2\rho_{12}\left(\frac{Y_{i1}-\mu_{1}}{\sigma_{1}}\right)\left(\frac{Y_{i2}-\mu_{2}}{\sigma_{2}}\right)+\left(\frac{Y_{i2}-\mu_{2}}{\sigma_{2}}\right)^{2}\right)\right\rbrace\\

\begin{multline}
l(\boldsymbol{\theta};\mathbf{Y}) =\prod_{i=1}^{n} \prod_{w=1}^{5}\prod_{k=1}^{2}f(Y_{i1},Y_{i2};\boldsymbol{\theta})\left[pr\left(Y_{i3}=w, Y_{i4}=k | Y_{i1}=y_{i1}, Y_{i2}=y_{i2}; \boldsymbol{\theta}\right)\right]^{I \lbrace Y_{i3}=w,Y_{i4}=k \rbrace}\label{obslike}
\end{multline}

The joint probability of patients having discrete measurements $Y_{i3}=w$ and $Y_{i4}=k$ must be multiplied over the five ordinal levels and two binary levels resulting in ten combinations of the probabilities in (\ref{bivprob}) to be calculated. We discuss the intuition for (\ref{bivprob}) in Appendix A. 

\begin{multline}\label{bivprob}
pr\left(Y_{i3}=w, Y_{i4} =k | Y_{i1}=Y_{i1}, Y_{i2}=Y_{i2}; \boldsymbol{\theta}\right)=\\\Phi_{2}\left(\tau_{w3}-\mu_{3|1,2}, \tau_{k4}-\mu_{4|1,2};\Sigma_{3,4|1,2}\right)-\Phi_{2}\left(\tau_{(w-1)3}-\mu_{3|1,2}, \tau_{k4}-\mu_{4|1,2};\Sigma_{3,4|1,2}\right)-\\\Phi_{2}\left(\tau_{w3}-\mu_{3|1,2}, \tau_{(k-1)4}-\mu_{4|1,2};\Sigma_{3,4|1,2}\right)+\Phi_{2}\left(\tau_{(w-1)3}-\mu_{3|1,2}, \tau_{(k-1)4}-\mu_{4|1,2};\Sigma_{3,4|1,2}\right)
\end{multline}

where $\Phi_{2}$ is the bivariate standard normal distribution function and $\mu_{3|1,2}, \mu_{4|1,2}$ and $\Sigma_{3,4|1,2}$ are derived using the rules of conditional multivariate normality, resulting in (\ref{condmean}).

\begin{equation}\label{condmean}
\begin{split}
\mu_{3|1,2}=&\mu_{3}+\tfrac{\left(\rho_{13}-\rho_{12}\rho_{23}\right)}{\sigma_{1}\left(1-\rho_{12}^{2}\right)}\left(Y_{i1}-\mu_{1}\right)+\tfrac{\left(\rho_{23}-\rho_{12}\rho_{13}\right)}{\sigma_{2}\left(1-\rho_{12}^{2}\right)}\left(Y_{i2}-\mu_{2}\right)\\
\mu_{4|1,2}=&\mu_{4}+\tfrac{\left(\rho_{14}-\rho_{12}\rho_{24}\right)}{\sigma_{1}\left(1-\rho_{12}^{2}\right)}\left(Y_{i1}-\mu_{1}\right)+\tfrac{\left(\rho_{24}-\rho_{12}\rho_{14}\right)}{\sigma_{2}\left(1-\rho_{12}^{2}\right)}\left(Y_{i2}-\mu_{2}\right)
\end{split}
\end{equation}

\begin{equation*}
\Sigma_{3,4|1,2}=\begin{pmatrix}
1-\frac{\rho_{13}^{2}-2\rho_{12}\rho_{13}\rho_{23}+\rho_{23}^{2}}{1-\rho_{12}^{2}} & \rho_{34}-\frac{\rho_{13}\rho_{14}-\rho_{12}\rho_{13}\rho_{24}-\rho_{12}\rho_{14}\rho_{23}+\rho_{23}\rho_{24}}{1-\rho_{12}^{2}} \\
\rho_{34}-\frac{\rho_{13}\rho_{14}-\rho_{12}\rho_{13}\rho_{24}-\rho_{12}\rho_{14}\rho_{23}+\rho_{23}\rho_{24}}{1-\rho_{12}^{2}}  &  1-\frac{\rho_{14}^{2}-2\rho_{12}\rho_{14}\rho_{24}+\rho_{24}^{2}}{1-\rho_{12}^{2}} 
\end{pmatrix}
\end{equation*}

\subsection{Estimation}

As the variance parameters $(\sigma_{1},\sigma_{2})$ are required to be greater than 0, we introduce parameters $(\delta_{1},\delta_{2})$ such that $\sigma_{1}=exp(\delta_{1})$ and $\sigma_{2}=exp(\delta_{2})$. This transformation ensures that the variance is above 0 whilst allowing the parameter we estimate to take any real value. We must also ensure that the correlation parameters $(\rho_{12},\rho_{13},\rho_{14},\rho_{23},\rho_{24},\rho_{34})$ are estimated within (-1,1) by introducing $ (\delta_{12},\delta_{13},\delta_{14},\delta_{23},\delta_{24},\delta_{34})$, where $\rho_{12}=2expit(\delta_{12})-1, \rho_{13}=2expit(\delta_{13})-1, \rho_{14}=2expit(\delta_{14})-1, \rho_{23}=2expit(\delta_{23})-1, \rho_{24}=2expit(\delta_{24})-1, \rho_{34}=2expit(\delta_{34})-1$.\\We fit the model in R by coding the likelihood function, probability of response and using the delta method to obtain standard errors. The bivariate distribution functions in (\ref{bivprob}) are estimated using `pmvnorm', using the method of \cite{Genz}.  The likelihood maximisation is conducted using the `nlminb' function in the `optimx' package, which is the best performing method in terms of accuracy and convergence rate, however is the slowest. We use the `Hessian' function in the `numDeriv' package to obtain the Hessian matrix and invert this to get the covariance matrix of the model parameters. In a small number of cases the Hessian is not positive definite because of computational error, meaning that it cannot be inverted. This is rectified in these cases by using the `near PD' function in the `Matrix' package, which computes the nearest positive definite matrix. 

\subsection{Inference}

We wish to make inference on the probability of response. Let $S_{i}$ be an indicator for patient i denoting whether or not they achieved response defined by $S_{i}$=1 if $ Y_{i1}^{}\leq\theta_{1}, Y_{i2}^{}\leq\theta_{2}, Y_{i3}^{*}\leq\theta_{3}, Y_{i4}^{*}\leq\theta_{4}$. Therefore,
\begin{equation}\label{probresp}
P(S_{i}=1\mid T_{i}, y_{i10}, y_{i20})=\int_{-\infty}^{\theta_{1}}\int_{-\infty}^{\theta_{2}}\int_{-\infty}^{\theta_{3}}\int_{-\infty}^{\theta_{4}} f_{\mathbf{Y}}(\mathbf{Y};T_{i},y_{i10},y_{i20})dy_{i4}^{*}dy_{i3}^{*}dy_{i2}^{}dy_{i1}^{}
\end{equation}

We obtain the integrand in (\ref{probresp}) by using the fitted values of the parameters in the conditional mean and conditional covariance matrix in (\ref{condmean}).  Parameter estimates from these methods are maximum likelihood estimates and so we avail of asymptotic maximum likelihood theory. The integral in (\ref{probresp}) is evaluated using the `R2Cuba' package to obtain estimates for each patient, assuming they were treated $\tilde{p_{i1}}$ and not treated $\tilde{p_{i0}}$. The odds ratio treatment effect is then defined as shown in (\ref{oddseffect}). 

\begin{equation}\label{oddseffect}
\tilde{\delta} = \dfrac{\left(\dfrac{\sum_{i=1}^{N}{\tilde{p_{i1}}}}{N-\sum_{i=1}^{N}{\tilde{p_{i1}}}}\right)}{\left(\dfrac{\sum_{i=1}^{N}{\tilde{p_{i0}}}}{N-\sum_{i=1}^{N}{\tilde{p_{i0}}}}\right)}
\end{equation}

Note that we can easily define a risk difference or risk ratio using these quantities but in what follows we consider $\tilde{\delta}$ to be the effect of interest. The standard error estimates are obtained using the delta method. This requires the covariance matrix of the maximum likelihood estimates Cov($\boldsymbol{\hat{\theta}}$) and $''\tilde{\boldsymbol{\delta}}$, the vector of partial derivatives of $\tilde{\delta}$ with respect to each of the parameter estimates. The variance of $\tilde{\delta}$ is obtained as shown in (\ref{deltaSE}).

\begin{equation}\label{deltaSE}
Var(\tilde{\delta})=(''\tilde{\boldsymbol{\delta}})^TCov(\boldsymbol{\hat{\theta}})(''\tilde{\boldsymbol{\delta}})
\end{equation}

Alternatively, the quantity in (\ref{probresp}) can also be considered to be a multivariate Gaussian hidden truncation distribution, from which we can obtain a closed form solution, and proceed as detailed by \cite{Arnold}. \\ Another important consideration for the model is how to assess goodness-of-fit. We propose an extension to an existing method for application in this case, which is detailed in Appendix B in the supplementary material. 

\section{Simulation study}
\label{sec3}
We are interested in comparing the performance of the latent variable, augmented binary and standard binary methods through simulation. The models for the augmented binary and standard binary methods are included in Appendix C in the supplementary material. 

\subsection{Data generating model}
Initially, we investigate the properties of the methods when the assumptions of the latent variable model are satisfied. The parameter values in the `baseline' scenario are chosen to simulate a scenario where composite endpoints are typically recommended for use. Namely, that all four components drive response and items are correlated but not so highly that the composite becomes redundant. The parameter values have been informed by the MUSE trial dataset, in particular the correlation structure. The response probability in the control arm is 0.275 and in the treatment arm is 0.381, resulting in an odds ratio equal to 1.6, values typically observed in trials requiring response in all four components. The parameter values selected for the model in (\ref{model1}) are shown in Table \ref{baselinesim}. From this baseline case, we vary parameters to determine how the methods behave under various scenarios of interest. In particular, under varying treatment effect, varying responder threshold and varying drivers of response. The parameter values for these data generating models are included in Appendix D in the supplementary material. 
\begin{table}[h!]
\caption{Parameter values for the data generating model in the baseline simulation scenario comparing the performance of the latent variable, augmented binary and standard binary methods for analysing a composite endpoint, where the values correspond to a treatment effect in all components and all components drive response}\label{baselinesim}
\begin{tabular}{ll}
\hline
\rule[-1ex]{0pt}{4ex}   Purpose & Values\\
\hline \\
Total sample size & N=300\\
Intercept & $\alpha_{0}=-4.9, \beta_{0}=-1.2, \psi_{0}=-0.2$\\
Treatment & $ \alpha_{1}=-0.28, \beta_{1}=-0.35, \gamma_{1}=-0.24, \psi_{1}=-0.18$\\
Baseline value & $\alpha_{2}=-0.5, \beta_{2}=-0.5$\\
Variance & $\sigma_{1}=\sigma_{2}=1$\\
Correlation & $\rho_{12}=0.5, \rho_{13}=\rho_{24}=0.35, \rho_{14}=0.25,\rho_{23}=0.4, \rho_{34}=0.3$\\
Discrete cut-point & $\tau_{13}=-1, \tau_{23}=-0.1, \tau_{33}=0.45, \tau_{43}=1.3$\\
Responder threshold & $\theta_{1}= -4, \theta_{2}=-0.6, \theta_{3}=0.45, \theta_{4}=0$\\
& \\
\hline
\end{tabular}
\end{table}
\subsection{Results}
The methods are evaluated against a range of performance criteria, which are included with their Monte Carlo standard errors in Appendix E of the supplementary material. For further details see \cite{Simstudy}.

\subsubsection{Varying treatment effect}
\begin{figure}[h!]
\centering \includegraphics[scale=0.9]{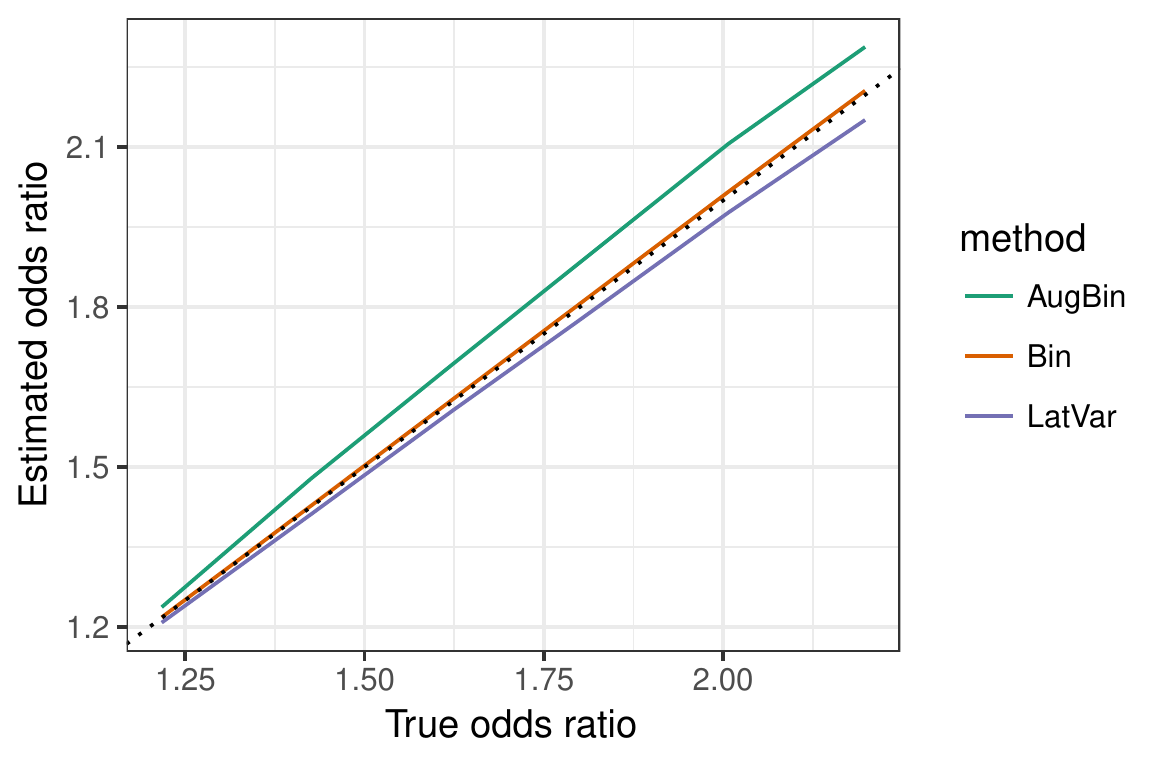}\caption{Bias reported from the latent variable method, augmented binary method and standard binary method when $n_{sim}$=5000, total sample size N=300 for true log-odds treatment effect between 1.2 and 2.2. The composite endpoint of interest contains four components: two continuous, one ordinal, one binary and treatment effects are present in all four components}\label{bias}
\end{figure}

\begin{figure}[h!]
\centering \includegraphics[scale=0.9]{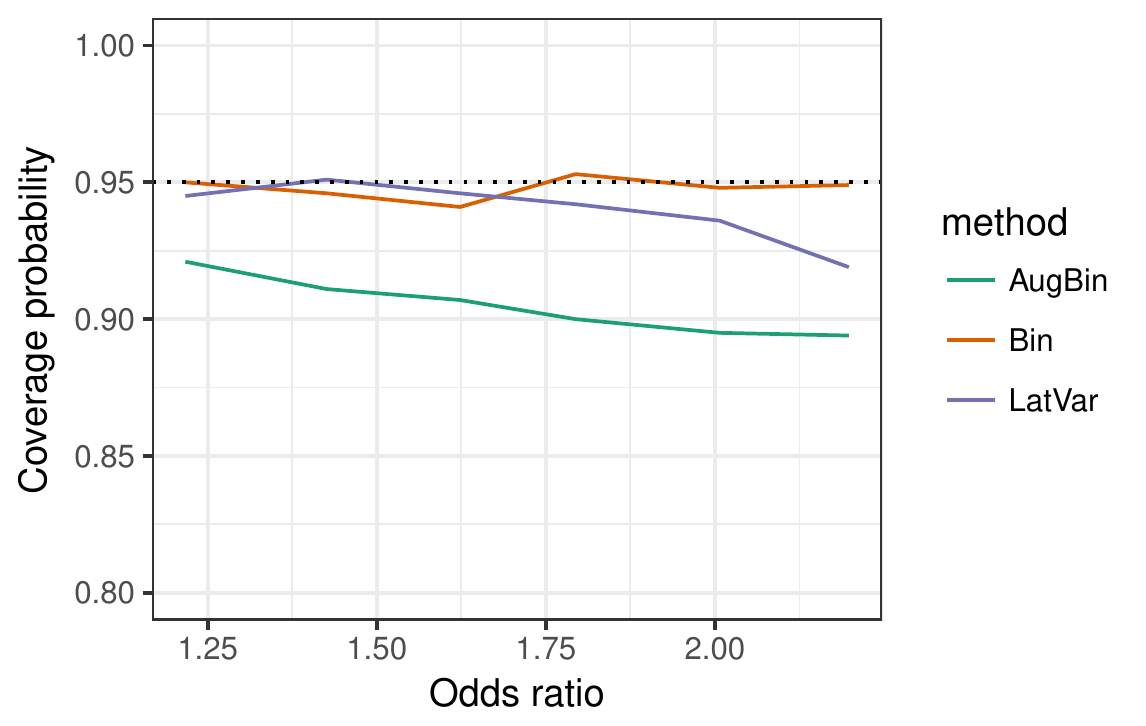}\caption{Coverage probability reported from the latent variable method, augmented binary method and standard binary method for $n_{sim}$=5000, total sample size N=300 for true log-odds treatment effect between 1.2 and 2.2. The composite endpoint of interest contains four components: two continuous, one ordinal, one binary and treatment effects are present in all four components}\label{coverage}
\end{figure}

\begin{figure}[h!]
\centering \includegraphics[scale=0.8]{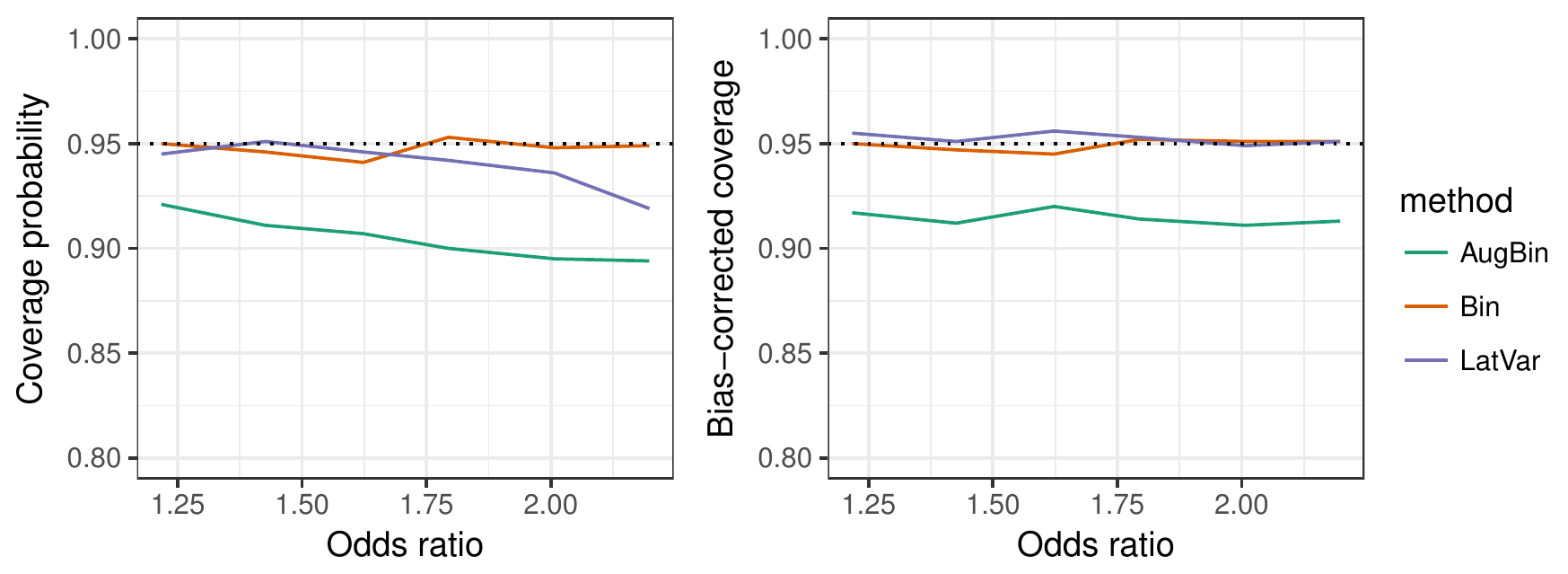}\caption{Coverage probability (left) and bias-corrected coverage probability (right) reported from the latent variable method, augmented binary method and standard binary method for $n_{sim}$=5000, total sample size N=300 for true log-odds treatment effect between 1.2 and 2.2. The composite endpoint of interest contains four components: two continuous, one ordinal, one binary and treatment effects are present in all four components}\label{covtreat}
\end{figure}

\begin{figure}[h!]
\centering \includegraphics[scale=0.9]{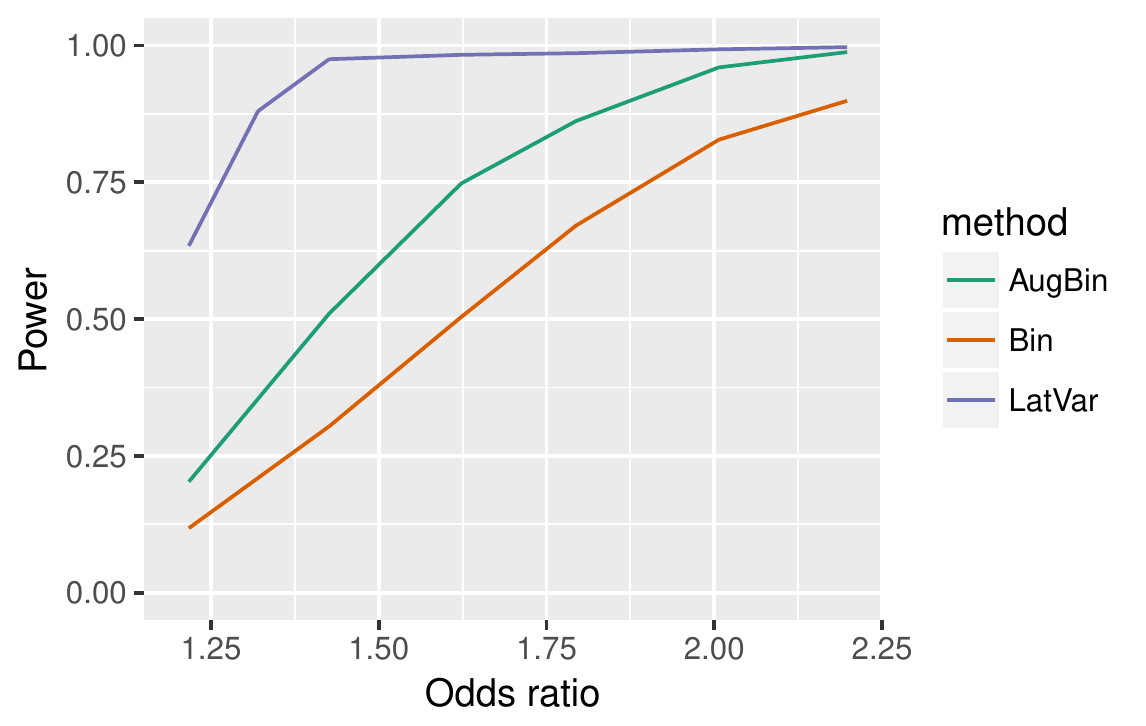}\caption{Statistical power reported from the latent variable method, augmented binary method and standard binary method for $n_{sim}$=5000, total sample size N=300 for true log-odds treatment effect between 1.2 and 2.2. The composite endpoint of interest contains four components: two continuous, one ordinal, one binary and treatment effects are present in all four components}\label{powertreat}
\end{figure}

\begin{figure}[h!]
\centering \includegraphics[scale=0.9]{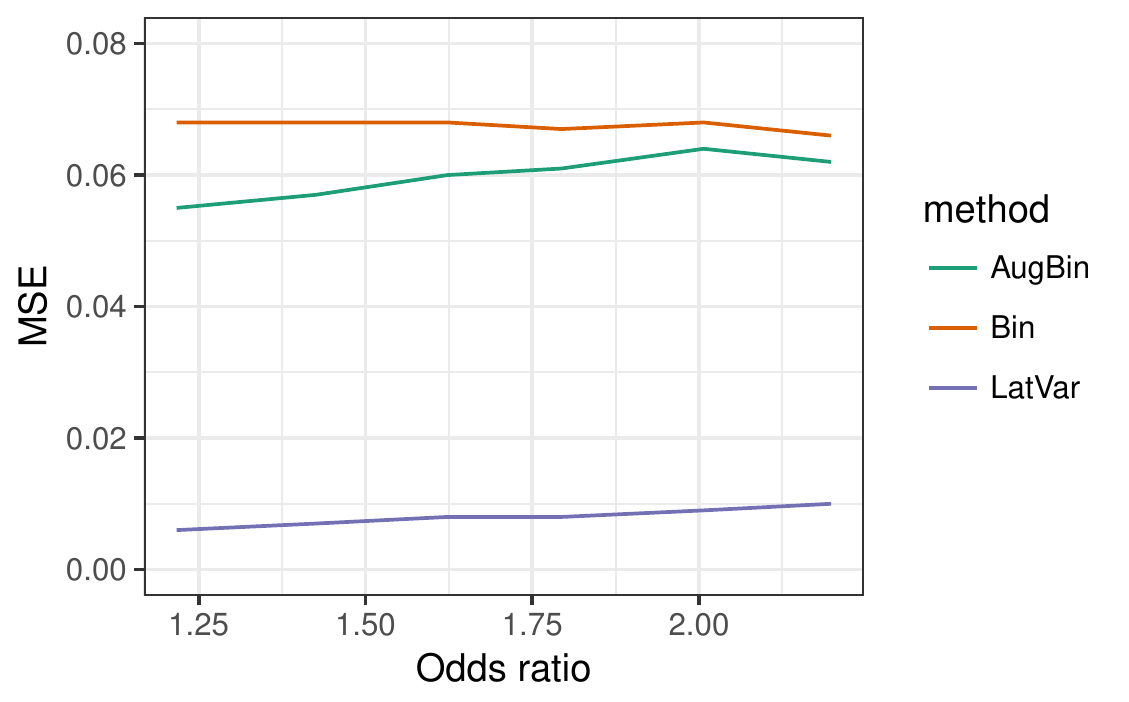}\caption{Mean Squared Error (MSE) reported from the latent variable method, augmented binary method and standard binary method for $n_{sim}$=5000, total sample size N=300 for true log-odds treatment effect between 1.2 and 2.2. The composite endpoint of interest contains four components: two continuous, one ordinal, one binary and treatment effects are present in all four components}\label{MSEtreat}
\end{figure}

Figure \ref{bias} shows the bias of the methods as the treatment effect varies. The standard binary method is unbiased, as we would expect for a logistic regression in a large sample. The latent variable method is unbiased for smaller treatment effects but a small bias towards the null is introduced as the treatment effect increases. The augmented binary method is biased away from the null in this setting and the bias increases as the treatment effect increases. Given that this performance is worse than is suggested from previous applications of the augmented binary method in \cite{WasonSeaman} and \cite{WasonJenkins}, this would suggest that the treatment effect from the augmented binary method may be biased if the model is misspecified. 
The coverage of the methods is shown in Figure \ref{coverage}. The binary method has approximately nominal coverage. The latent variable method has nominal coverage for smaller treatment effects, however the coverage probability decreases as the treatment effect increases. The augmented binary method has coverage of approximately 0.91, which also decreases when the treatment effect increases. 
In order to diagnose this under-coverage in the joint modelling methods we can look at bias-corrected coverage, as recommended in \cite{Simstudy}. Figure \ref{covtreat} shows both the coverage and bias-corrected coverage for the three methods. The properties of the standard binary method remain unchanged. The bias-corrected coverage of the latent variable method is 0.95, which indicates that any under-coverage is due to the bias present. This is not true for the augmented binary method which shows small improvements in bias-corrected coverage, indicating that under-coverage is present in this method due to reasons other than bias. Again, this may be down to model misspecification.   
The power of the three methods is shown in Figure \ref{powertreat}. The performance of the binary and augmented binary methods are as we would expect based on previous findings in \cite{WasonSeaman} and \cite{WasonJenkins}. The latent variable method offers much higher power. In this setting it has close to 100\% power for odds ratios larger than 1.6, an effect that is plausible to observe in a trial. 
These findings have indicated that the standard binary method has the smallest bias and that the latent variable method has the smallest variance. The mean squared error (MSE) provides a combined measure of bias and variance. Figure \ref{MSEtreat} shows the MSE of the three methods as the treatment effect varies. The MSE for the standard and augmented binary methods is approximately 6.5 times that of the latent variable method. However, this measure should be interpreted with care due to the fact that the MSE is more sensitive to the sample size than comparisons of
bias or empirical SE alone (\cite{Simstudy}). 

\subsubsection{Varying $\theta_{1}$}

\begin{figure}[h!]
\centering \includegraphics[scale=1]{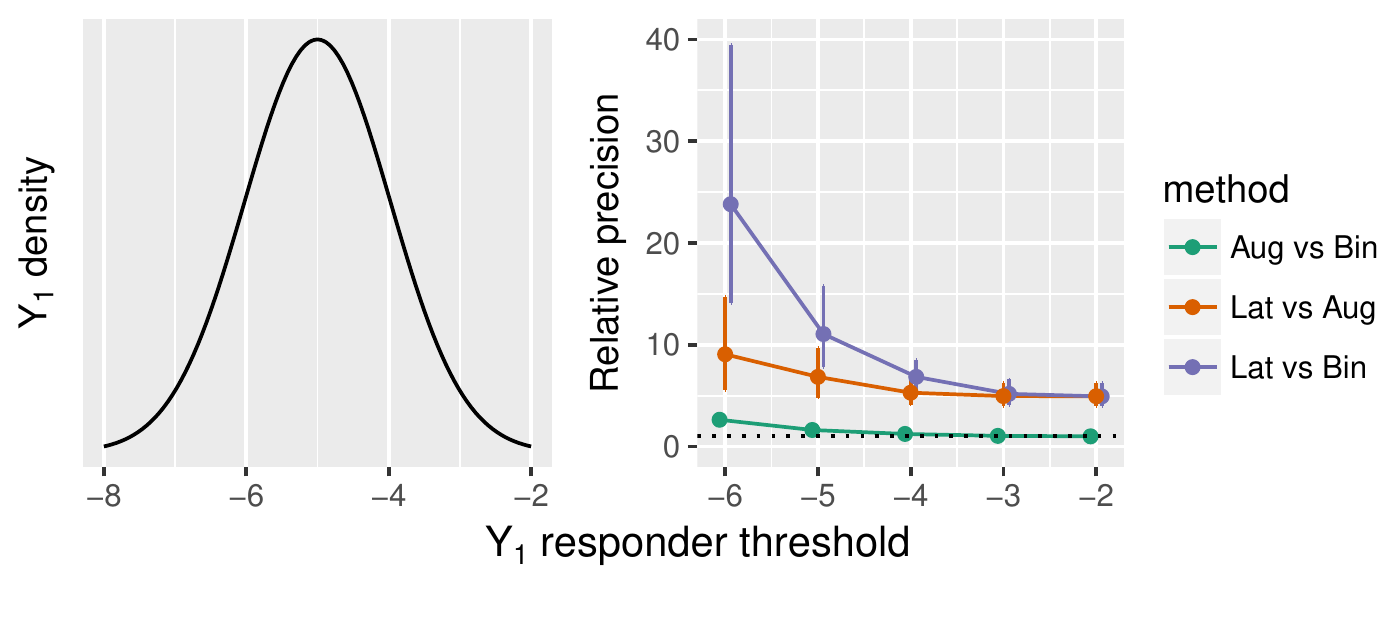}\caption{Density of continuous $Y_{1}$ variable (left) and estimated relative precision of augmented binary versus standard binary method, latent variable versus augmented binary method and latent variable versus standard binary method as the $Y_{1}$ responder threshold $\theta_{1}$ varies between $\theta_{1}=-6$ and $\theta_{1}=-2$ (right) for $n_{sim}$=5000 and total sample size N=300. The composite endpoint of interest contains four components: two continuous, one ordinal, one binary and treatment effects are present in all four components}\label{Y1thres}
\end{figure}

To understand more about the precision performance of the augmented binary method in particular, we vary the responder threshold $\theta_{1}$ to change the proportion of responders in that outcome. Figure \ref{Y1thres} shows the density of the $Y_{1}$ variable and the relative precision of the methods, as the responder threshold varies. The precision gains from the augmented binary method diminish as the threshold increases. This is intuitive, as improvements in efficiency fall as the continuous component becomes less responsible for driving response. It is interesting to note that all precision gains are lost for any thresholds above -4. Therefore, even when 20\% of patients are non-responders, all efficiency gains are lost. The percentage of responders needed to improve efficiency using the augmented binary method will of course depend on the correlation structure employed. Due to the additional information in the other components, the latent variable method is still five times as precise as the other methods. 

\subsubsection{Components contributing to response}

\begin{figure}[h!]
\centering \includegraphics[scale=1]{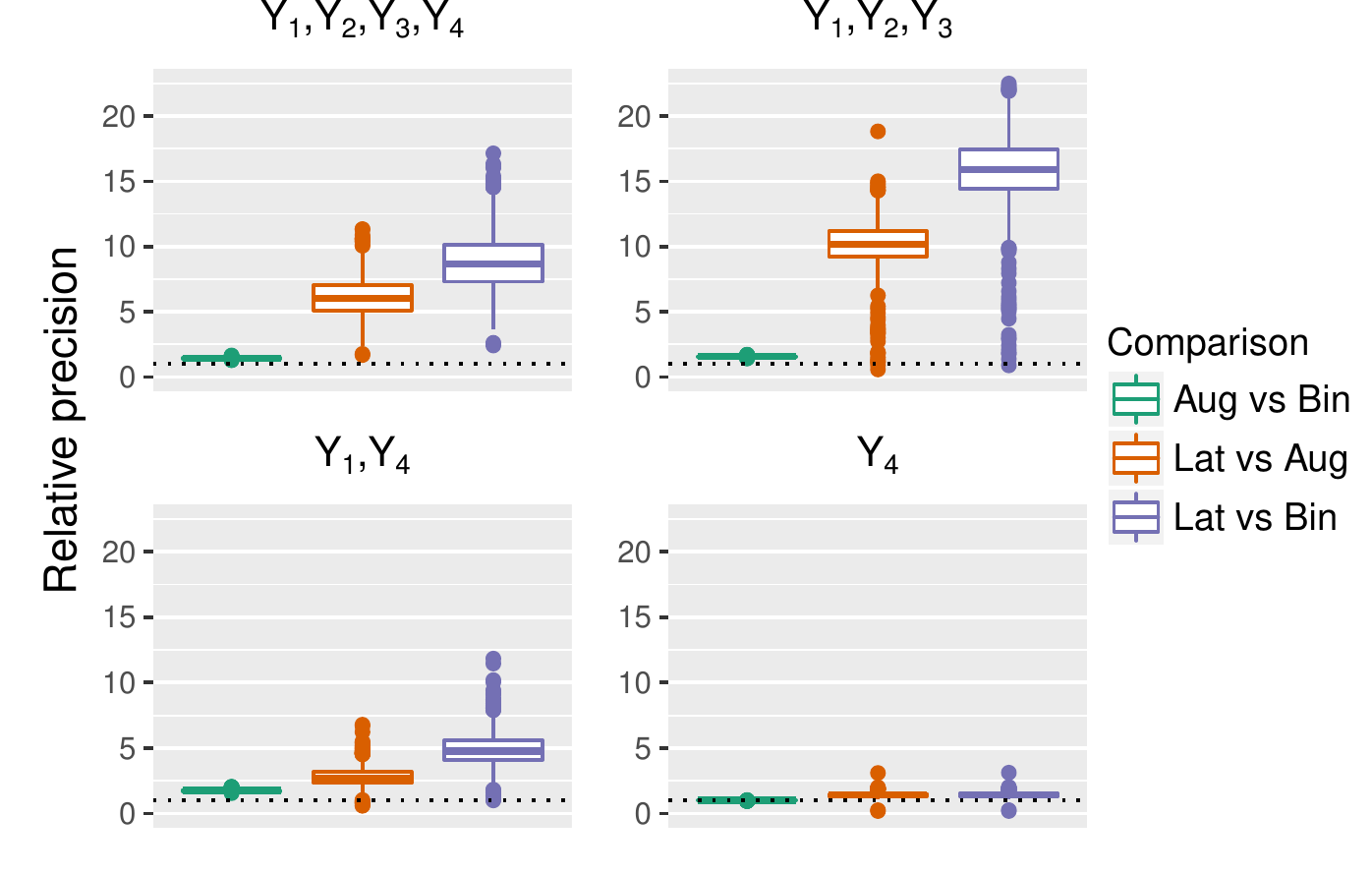}\caption{Estimated relative precision gains from augmented binary versus standard binary method, latent variable versus augmented binary method and latent variable versus standard binary method when different combinations of components driving response. Response driven by $Y_{1},Y_{2},Y_{3},Y_{4}$ (top left), $Y_{1},Y_{2},Y_{3}$ (top right), $Y_{1},Y_{4}$ (bottom left) and $Y_{4}$ (bottom right) where $Y_{1},Y_{2}$ are continuous, $Y_{3}$ is ordinal, $Y_{4}$ is binary for $n_{sim}$=5000 and total sample size N=300. The composite endpoint of interest contains four components: two continuous, one ordinal, one binary and treatment effects are present in all four components}\label{precthres}
\end{figure}

An important consideration when investigating performance is how the precision changes when different combinations of outcomes are responsible for driving response. Figure \ref{precthres} shows boxplots of the relative precision for the methods for four different response combinations, namely when response is driven by $(Y_{1}, Y_{2}, Y_{3}, Y_{4}), (Y_{1}, Y_{2}, Y_{3}), (Y_{1}, Y_{4})$ and $(Y_{4})$, where $Y_{1}$ and $Y_{2}$ are observed as continuous variables, $Y_{3}$ is ordinal and $Y_{4}$ is binary.\\
When all four components contribute to response, the latent variable method outperforms the other methods, offering large precision gains. The variability in the magnitude of these gains is large, with the median result showing that the latent variable method reports the treatment effect 8 times more precisely than the binary method and 6 times more precisely than the augmented binary method. If response is driven by $(Y_{1},Y_{2},Y_{3})$ then the relative median gains for the latent variable method are larger, however note that in less that 2\% of cases the treatment effect is reported equally or less precisely than from both of the other methods. The findings are similar when response is driven by $(Y_{1},Y_{4})$, however the median gains are much smaller. The treatment effect is reported 5 times more precisely from the latent variable method than the binary method in this setting. Note that as the augmented binary method models the relevant components it still performs well and again better than the latent variable method in a very small number of cases. When binary $Y_{4}$ determines response, the augmented binary method offers no improvement in precision whereas the latent variable method is approximately 1.5 times more precise. It is clear from the results that the magnitude of the precision gains from the latent variable method is highly dependent on the structure of the data.

\subsection{Sensitivity analysis}
The key assumptions in this model are that of joint normality of the four components and that the discrete variables are realisations of latent continuous variables. Although it is not possible to test these assumptions in real data, we can investigate how robust the latent variable method is to deviations from these conditions. We can do this by drawing from the multivariate skew-normal distribution with different degrees of skew in each of the components. The first scenario investigated considers when all four components are skewed.  Scenarios 2-3 consider different magnitudes of skew in the latent continuous components only. This tests the robustness of the method to the assumption that the observed discrete variables manifest from a true normal continuous variable. Scenario 4 is the null case for scenario 3. The results are shown in Appendix F of the supplementary material. 

In summary, scenarios 1-3 have increased bias resulting in under-coverage as the bias-corrected coverage is close to nominal for all scenarios. The coverage of the latent variable method is nominal in the null case. This is consistent with our previous findings however the magnitude of the bias is much smaller when the assumptions are satisfied. The latent variable method still offers large power gains over the other methods. The MSE is smallest for the latent variable method across all scenarios investigated, indicating that the large reduction in variance is useful despite the introduction of bias. The latent variable method estimates the probability of response in the control arm well however underestimates the probability of response in the treatment arm. The magnitude of this underestimation is unaffected by the degree of skew or whether the skew is present in the observed continuous components. The relative precision of the methods are consistent with our previous findings indicating that the violation of joint normality only affects the bias and not the variance. The augmented binary and standard binary methods behave similarly to when the joint normality assumptions are satisfied, which is expected given that the assumptions of those models are violated in both contexts.

\section{Case study}
\label{sec4}

\subsection{Data structure}
\begin{figure}[h!]
\centering \includegraphics[scale=0.9]{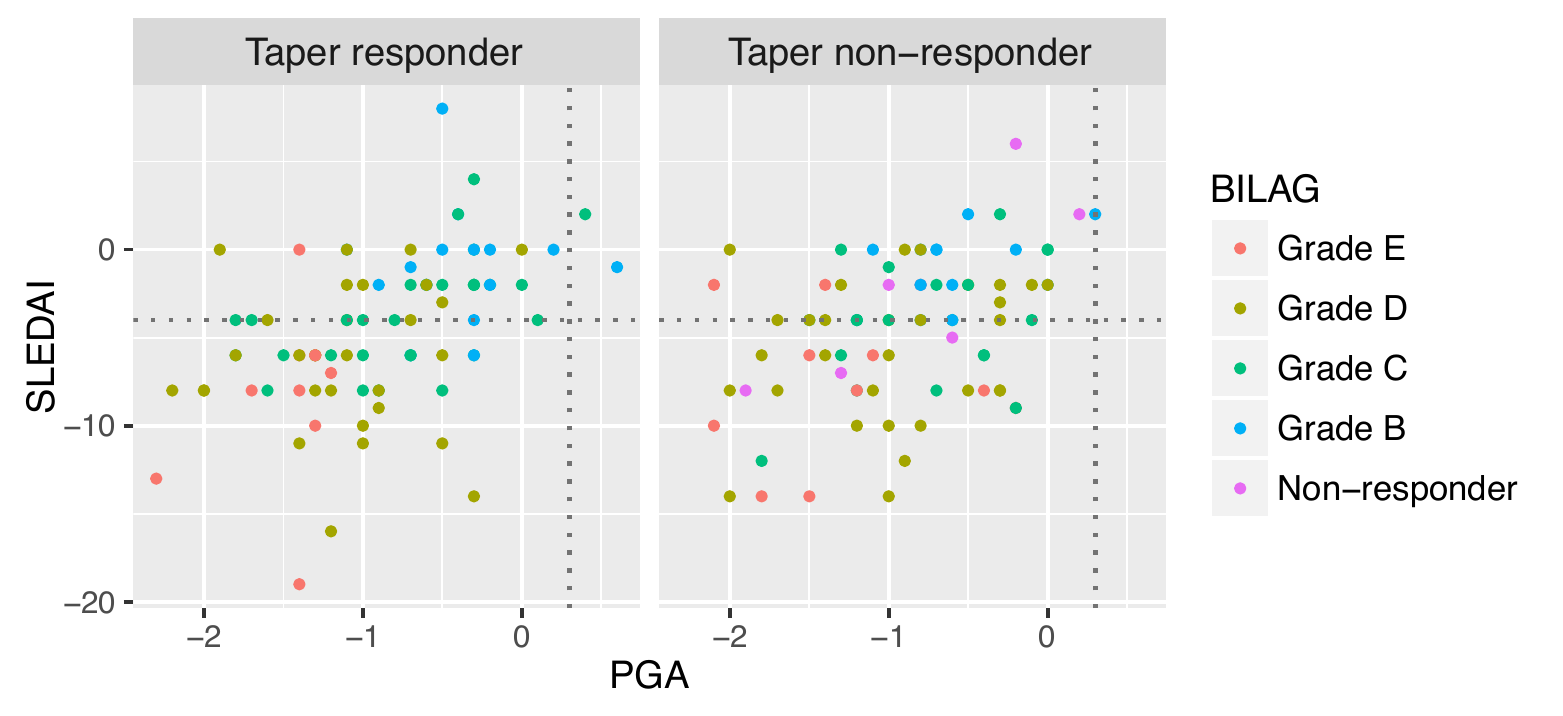}\caption{Observed response rates in each of the SLE responder index components in the Phase IIb MUSE trial. SLEDAI is plotted on the y-axis and PGA on the x-axis, along with their corresponding dichotomisation thresholds. Levels of BILAG are represented by different colours and taper responders and non-responders are split across two panels. }\label{responderplot}
\end{figure}

Due to data sharing policy, we conduct the analysis for a subset of the patients, N=278 rather than N=305 reported in the paper, so the results will differ from the original paper. Furthermore, only the anifrolumab 300mg arm (n=95) and the placebo arm (n=87) will be used to illustrate the methods.\\ The simulation results have suggested that the structure of the data is important for how the methods will perform, in particular the magnitude of the precision gains depends highly on which components drive response. Table \ref{components} shows the criteria for response in each component and the rates of response in each by treatment arm. This suggests that the components responsible for responder discrimination are the continuous SLEDAI measurement and the binary taper measure. 
 \begin{table}[h!]
\caption{Observed response rates in each of the SLE responder index components in the anifrolmab 300mg arm and placebo arm of the Phase IIb MUSE trial. SLE index is comprised of a continuous SLEDAI outcome, continuous PGA outcome, ordinal BILAG outcome and binary taper outcome where response in each component is achieved when the patient meets the criteria shown}\label{components}
\centering
\begin{tabular}{cccc}
\hline
\rule[-1ex]{0pt}{4ex}  Components& Response criteria & \multicolumn{2}{c}{Treatment arm}\\
\cmidrule(r){3-4}
\rule[-1ex]{0pt}{4ex}   & & Anifrolumab 300mg &  Placebo \\
\hline
&&&\\
SLEDAI  & Change in SLEDAI $\leq$ -4 & 58/89 & 41/76\\
&&&\\
PGA & Change in PGA $<$ 0.3& 87/89 & 75/76\\
& & &\\
BILAG & No Grade A or more than & 86/89& 72/76\\
&one Grade B&&\\
& & &\\
Taper&Sustained reduction in & 53/95 & 37/87\\
&oral corticosteroids&&\\
&&&\\
SLE responder index& Responder in all four components & 34/95 &18/87\\
&&&\\
\hline
\end{tabular}
\end{table}
We can further explore the structure of the data by visualising the 4-D endpoint. Figure \ref{responderplot} shows a plot of the four components in the SLE index. The two panels show taper responders and non-responders, the levels in BILAG are denoted using colours where any coloured data points representing Grade B - Grade E are responders. The response thresholds for the continuous measurements are included, where a patient must be below the threshold to be considered a responder. We can conclude that response is entirely driven by SLEDAI and the taper variable, as there are no PGA non-responders not already accounted for by SLEDAI and no purple data points in the responder quadrant. 
\subsection{Results}

The probability of response in the placebo arm is estimated as 0.199 by the latent variable method, 0.211 by the augmented binary method and 0.224 by the standard binary method. A much larger discrepancy between the methods is shown in the treatment arm, where the probability of response is estimated at 0.311, 0.324 and 0.382 in the latent variable, augmented binary and standard binary methods respectively.

The log-odds treatment effect point estimates and confidence intervals for the MUSE trial are shown in Table \ref{bootstrap}. Both joint modelling methods estimate the treatment effect more precisely. Although there may be bias present in the point estimates for the joint modelling methods, the confidence intervals entirely overlap with that of the binary method. All three methods indicate that anifrolumab 300mg performs better than placebo, as in the original findings. The latent variable model fits the data well according to the modified Pearson residuals, see Appendix G. 

The simulation results indicated that the latent variable method may report the treatment effect with bias and have problems with bias related under-coverage when the treatment effect is large and when the assumption of joint normality is not satisfied. As the problems with the performance are bias related, we suggest implementing a bootstrap procedure to correct for this. In this scenario N=182 and $n_{boot}$=1000, therefore the procedure is as follows:
\begin{enumerate}
\item Sample with replacement N=182 patients from the MUSE trial 
\item Compute the treatment effect using the latent variable, augmented binary and standard binary methods
\item Repeat step 1 and 2 $n_{boot}$=1000 times
\item Obtain an estimate of the bias using the difference between the treatment effect in the MUSE trial and the mean of the bootstrap treatment effects 
\end{enumerate}

A 95\% bootstrap confidence interval for the treatment effect estimate can be obtained by ordering the 1000 bootstrap estimates of the treatment effect and taking the $25^{th}$ and $975^{th}$ estimate. The point estimates and 95\% confidence intervals from the MUSE trial and from the bootstrap re-sampling are shown in Table \ref{bootstrap}.
\begin{table}[h!]
\caption{Log-odds treatment effect estimates and 95\% confidence intervals from the latent variable method, augmented binary method and standard binary method in the Phase IIb MUSE trial and the bootstrap sample when N=182 and $n_{boot}=1000$}
\label{bootstrap}
\centering
\begin{tabular}{ccccc}
\hline
\rule[-1ex]{0pt}{4ex}  Method &\multicolumn{2}{c}{Log-odds treatment effect}\\
 \cmidrule(r){2-3}
 \rule[-1ex]{0pt}{4ex}  &  MUSE trial estimate & Bootstrap estimate & \\  \hline
 &&&\\
\hspace{0.6cm} Latent Variable & 0.641 (0.217, 1.072)& 0.682 (0.275, 1.137) \\
&&&\\
\hspace{0.6cm} Augmented binary & 0.580 (0.139, 1.021) & 0.608 (0.096, 1.111) \\
&&&\\
\hspace{0.6cm} Binary &  0.763 (0.078, 1.449) & 0.809 (0.112, 1.561)\\
&&&\\
\hline
\end{tabular}\\
\end{table}
The log-odds point estimate from the latent variable method has been shifted away from the null by approximately 0.04. This is the magnitude of bias that the simulation results suggested for this treatment effect. The width of the confidence interval hasn't changed much from the original estimate in the bootstrap sample, indicating that the variance is well estimated in the trial dataset. The point estimate for the binary method has also been shifted substantially, despite the simulations showing this method to be unbiased. This is likely due to the large imprecision in the treatment effect reported by the binary method.

In terms of estimated precision, it is interesting to determine where the trial data set lies in the distribution of datasets generated in the simulation study. The latent variable method reports the treatment effect 2.5 times more precisely than the standard binary method in this setting, whilst the augmented binary method is 2.4 times more precise. We would have expected this similar performance as the augmented binary method models the SLEDAI and taper variables - the only components driving response. This increase in precision from the latent variable method compared with the binary method amounts to a 60\% reduction in required sample size.

\section{Discussion}
\label{sec5}

In this paper we addressed the issue of substantial losses of information when modelling complex composite endpoints. By employing concepts of partitioning latent variable outcome spaces we could model the observed structure of the composite endpoint, which resulted in large gains in efficiency. Sensitivity analyses showed that a bias is introduced when the assumptions of joint normality were not satisfied, however similar reductions in variance were observed. When applying the methods to the MUSE trial, we implemented a bootstrap procedure to correct for the presumed bias, as joint normality could not be assessed. The treatment effect was reported 2.5 times more precisely than that reported from the standard binary method. \\
Bias correction appears to perform well in the real data, where the crucial assumptions cannot be tested. The point estimate is shifted by a magnitude that would have been expected from the simulation results and the estimate of the variance is similar to that obtained in the single trial dataset. Furthermore the bootstrap confidence interval for the treatment effect is contained within that for the binary method, which offers further reassurance for application. However, more work could be done to investigate different structures and scenarios to ensure that the bias correction is always performing as expected. Ideally, we would investigate this further across a large number of datasets however this is too computationally intensive. To perform this on one replicate, where $n_{boot}=1000$ using 200 cores on a high performance computer (HPC) currently takes 7 hours. Exploring this further through bootstrapping or employing alternative multivariate distributions is an area for future research.\\
The precision gains offered by the latent variable method offer justification for the additional complexity. However, the magnitude of these gains are highly dependent on the components that drive response. The baseline case in the simulations was chosen to reflect when a composite endpoint is recommended for use, i.e. when all four components determine response rates. In this scenario, the precision gains achieved resulted in the latent variable method reporting the effect 2.5 to 17.5 times more precisely than the standard binary method. However, in practice in SLE trials, this has not been found to be the case. A review of two phase 3 trials (N= 2262) using the SRI-5 index found the SRI-5 response rate at week 52 for all patients was 32.8\% (\cite{SRIdrivers}). Non-response due to a lack of SLEDAI improvement, concomitant medication non-compliance or dropout was 31, 16.5 and 19.1\%, respectively. Non-response due to deterioration in BILAG or Physician’s Global Assessment after SLEDAI improvement, concomitant medication compliance and trial completion was 0.5\%. This is in agreement with our findings from the MUSE trial data, which suggests that the precision gains in the baseline case are optimistic. The simulation results show that when one continuous and one binary component drive response, the latent variable method may be anywhere between 1 and 12 times as precise as the binary method and up to 7 times as precise as the augmented binary method. In a very small number of cases ($<$2\%) there are no efficiency gains from using the latent variable method in this scenario. However the potential gains available in 98\% of cases ensures that implementing the latent variable method is still very much a worthwhile endeavour, for all stakeholders in a clinical trial. \\
In addition to SLE, we have identified other disease areas that have a similar complex composite structure, meaning the potential to improve efficiency extends well beyond SLE. However, it must be acknowledged that the exact structure of the endpoint may offer different magnitudes of bias and precision, and may require longer computational time. Furthermore, in conditions where longitudinal data is required to sufficiently capture disease activity, trials may include multiple follow-up times and the method will need to be extended to include latent variables in the mean structure to account for this. In terms of scalability to more complex endpoints, the computational time depends on many things, in particular the number of outcomes, the outcome scale and the number of levels in the ordinal variable. In our case, we find the number of ordinal levels to be the most influential factor in computational time. This is due to the fact that 5 levels in the ordinal variable leads to 10 probability calculations in (\ref{bivprob}), however 3 levels would require the computation of 6 joint probabilities. Consequently, the run time will be substantially increased if there are multiple ordinal levels and decreased if the discrete variables are binary. If the computational time for a particular endpoint is deemed to be too large, then we may reduce the complexity of the endpoint by collapsing the least informative components in to a single binary variable. It must be acknowledged that as we have coded the likelihood, with no package available to do this, the likelihood and probability of response code will have to be tailored specifically to each endpoint. The potential gains in efficiency justify this additional complexity.\\
We have shown that the latent variable method is a powerful tool in composite endpoint analysis and should be considered as a primary analysis method in a trial using these endpoints. In order for implementation in the general case, where the composite contains any number of continuous and discrete outcomes and to ensure the uptake of the method in clinical trials, we will need to develop a software package. Furthermore, if patients and investigators are to benefit from the efficiency gains, we will need a method to calculate the required sample size in a given trial. We are currently working on addressing these issues to aid in the application of the method.

\section*{Acknowledgments}
{\it Disclosure}: The MUSE trial data set was received under a data sharing contract with AstraZeneca. 

\bibliographystyle{biorefs}
\bibliography{bibliography}

\section*{Supplementary material}
\section*{Appendix A}

The joint probability below expresses, for patient i with $Y_{1}=y_{i1}$ and $Y_{2}=y_{i2}$, the probability that they will have a $Y_{3}$ score $w$ and a $Y_{4}$ score $k$. 
\begin{multline}
pr\left(Y_{i3}=w, Y_{i4} =k | Y_{i1}=Y_{i1}, y_{i2}=y_{i2}; \boldsymbol{\theta}\right)=\\\Phi_{2}\left(\tau_{w3}-\mu_{3|1,2}, \tau_{k4}-\mu_{4|1,2};\Sigma_{3,4|1,2}\right)-\Phi_{2}\left(\tau_{(w-1)3}-\mu_{3|1,2}, \tau_{k4}-\mu_{4|1,2};\Sigma_{3,4|1,2}\right)-\\\Phi_{2}\left(\tau_{w3}-\mu_{3|1,2}, \tau_{(k-1)4}-\mu_{4|1,2};\Sigma_{3,4|1,2}\right)+\Phi_{2}\left(\tau_{(w-1)3}-\mu_{3|1,2}, \tau_{(k-1)4}-\mu_{4|1,2};\Sigma_{3,4|1,2}\right) \tag{A.1}
\end{multline}

The intuition for the joint probability can be seen below in Figure \ref{bivprobfig}, specifically for the SLE endpoint, where $w=5$ and $k=2$.\\ 

\begin{figure}[h!]
\centering \includegraphics[scale=0.85]{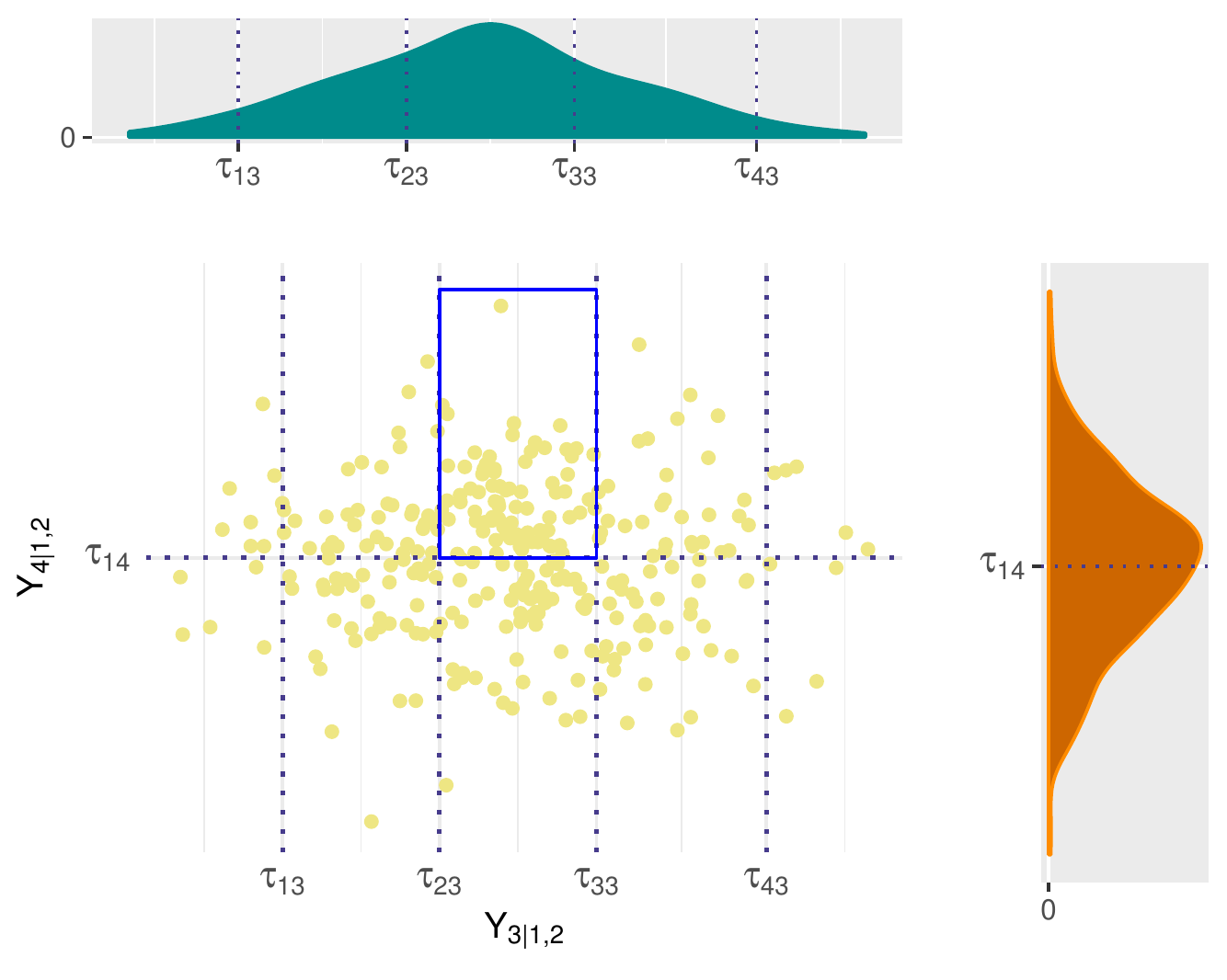}\caption{The figure shows the conditional outcome $Y_{3|1,2}$ on the x-axis and $Y_{4|1,2}$ on the y-axis with their corresponding underlying continuous densities and partitioning thresholds. The area where $w=3$ and $k=2$ is highlighted for illustration}\label{bivprobfig}
\end{figure}
\vspace{5mm}
The blue box indicates the region where $w=3$ and $k=2$. As $\tau_{03}=\tau_{04}=-\infty$ and $\tau_{53}=\tau_{24}=\infty$, the corresponding probability is shown in (\ref{bivprobeg}).
\begin{multline}\label{bivprobeg}
pr\left(Y_{i3}=3, Y_{i4} =2 | Y_{i1}=y_{i1}, Y_{i2}=y_{i2}; \boldsymbol{\theta}\right)=\\\Phi_{2}\left(\tau_{33}-\mu_{3|1,2}, \infty-\mu_{4|1,2};\Sigma_{3,4|1,2}\right)-\Phi_{2}\left(\tau_{23}-\mu_{3|1,2}, \infty-\mu_{4|1,2};\Sigma_{3,4|1,2}\right)-\\\Phi_{2}\left(\tau_{33}-\mu_{3|1,2}, \tau_{(14}-\mu_{4|1,2};\Sigma_{3,4|1,2}\right)+\Phi_{2}\left(\tau_{23}-\mu_{3|1,2}, \tau_{14}-\mu_{4|1,2};\Sigma_{3,4|1,2}\right) \tag{A.2}
\end{multline}

 \clearpage

\section*{Appendix B}

One suggestion in the literature for assessing goodness-of-fit in latent variable models is introduced by \cite{ordinal3} for the case when there is one continuous and one ordinal variable. This may be extended to allow for two continuous, one ordinal and one binary outcome for application in SLE, as shown below. \\
As before, let $\mathbf{Y}_{i}=(Y_{i1}, Y_{i2}, Y_{i3}, Y_{i4})'$ be the vector of observed responses for patient i. Then, partitioning the observed and latent continuous measures, we let $\mathbf{Y_{cts}}=(Y_{1}^{}, Y_{2}^{})$ and $\mathbf{Y_{dis}}=(Y_{3}, Y_{4})$. Then, $\hat{\Sigma}_{11}=\hat{Var}(\mathbf{Y_{cts}}), \hat{\Sigma}_{22}=\hat{Var}(\mathbf{Y_{dis}}), \hat{\Sigma}_{12}=\hat{\Sigma}_{21}=\hat{Cov}(\mathbf{Y_{cts}},\mathbf{Y_{dis}})$.\\The modified Pearson residuals, taking in to account the correlation between responses are shown below. \begin{equation}\tag{B.1}
r_{i}^{p} = \hat{\Sigma}^{-\frac{1}{2}}(Y_{i}-\hat{\mu_{i}})\label{modpearson}
\end{equation} where, \begin{equation}\tag{B.2}
\boldsymbol{\hat{\mu_{i}}}=(\hat{E}(Y_{i1}, Y_{i2}\mid X_{i1}, X_{i2}),\hat{E}(Y_{i3}, Y_{i4}\mid X_{i3}, X_{i4}))'
\end{equation} and \begin{equation}\tag{B.3}
\hat{\Sigma}=\begin{pmatrix}
\hat{\Sigma}_{11} & \hat{\Sigma}_{12}  \\
\hat{\Sigma}_{21}  & \hat{\Sigma}_{22}  \\
\end{pmatrix}
\end{equation}

\vspace{5mm}

A Cholesky decomposition may be used to obtain $\hat{\Sigma}^{-\frac{1}{2}}$ in (\ref{modpearson}). The covariance between the vector of observed continuous and observed discrete responses is shown below. \begin{align*}
\Sigma_{12}= & E(\mathbf{Y_{cts}Y_{dis}}) - E(\mathbf{Y_{cts}})E(\mathbf{Y_{dis}}) \\
= & E(\mathbf{Y_{cts}} E(\mathbf{Y_{dis}} \mid \mathbf{Y_{cts}})) - E(\mathbf{Y_{cts}})E(\mathbf{Y_{dis}}) \\
= & E(Y_{1}Y_{2}E(Y_{3}, Y_{4} \mid Y_{1}, Y_{2})) - E(\mathbf{Y_{cts}})E(\mathbf{Y_{dis}})\\
= & \int_{y_{1}}\int_{y_{2}} y_{1}y_{2} \sum_{y_{3}} \sum_{y_{4}}y_{3}y_{4} P\left(Y_{3}=w, Y_{4} =k | Y_{1}=y_{1}, Y_{2}=y_{2}\right)f_{Y_{1},Y_{2}}(y_{1}, y_{2})dy_{1}dy_{2} \\
& \hspace{93mm} - E(\mathbf{Y_{cts}})E(\mathbf{Y_{dis}})\\
\end{align*} Where,
\begin{multline*}
P\left(Y_{3}=w, Y_{4} =k | Y_{1}=y_{1}, Y_{2}=y_{2}\right)=\\\Phi\left(\tau_{w3}-\mu_{3|1,2}, \tau_{k4}-\mu_{4|1,2};\Sigma_{3,4|1,2}\right)-\Phi\left(\tau_{(w-1)3}-\mu_{3|1,2}, \tau_{k4}-\mu_{4|1,2};\Sigma_{3,4|1,2}\right)-\\\Phi\left(\tau_{w3}-\mu_{3|1,2}, \tau_{(k-1)4}-\mu_{4|1,2};\Sigma_{3,4|1,2}\right)+\Phi\left(\tau_{(w-1)3}-\mu_{3|1,2}, \tau_{(k-1)4}-\mu_{4|1,2};\Sigma_{3,4|1,2}\right)
\end{multline*}
 \begin{equation*}
E(\mathbf{Y_{cts}})=\int_{y_{1}}\int_{y_{2}} y_{1}y_{2} f_{Y_{1},Y_{2}}(y_{1}, y_{2})dy_{1}dy_{2}
\end{equation*} \begin{equation*}
E(\mathbf{Y_{dis}})=\sum_{y_{3}} \sum_{y_{4}}y_{3}y_{4} P\left(Y_{3}=w, Y_{4} =k \right)
\end{equation*} and \begin{align*}
P(Y_{3}=w, Y_{4} =k)= \Phi(&\tau_{w3}-\mu_{3}, \tau_{k4}-\mu_{4};\rho_{3,4})-\Phi(\tau_{(w-1)3}-\mu_{3}, \tau_{k4}-\mu_{4};\rho_{3,4})-\\ \Phi(&\tau_{w3}-\mu_{3}, \tau_{(k-1)4}-\mu_{4};\rho_{3,4})+\Phi(\tau_{(w-1)3}-\mu_{3}, \tau_{(k-1)4}-\mu_{4};\rho_{3,4})
\end{align*}\label{margbivprob} The Pearson residual is based on the Pearson goodness-of-fit statistics \begin{equation}\tag{B.4}
\chi_{p}^{2}=\sum_{i=1}^{n} \chi_{p}^{2} (Y_{i}, \hat{\mu_{i}})
\end{equation} with ith component \begin{equation}\tag{B.5}
\chi_{p}^{2} (Y_{i}, \hat{\mu_{i}}) = (Y_{i}-\hat{\mu_{i}})'\hat{\Sigma}^{-1}(Y_{i}-\hat{\mu_{i}})
\end{equation} The distribution of the residuals should follow a chi-squared distribution with p degrees of freedom. Comparing the residuals to the chi-squared value allows us to identify observations which the model does not fit well. If there are many observations unexplained by the model then it could indicate a poor choice of model. This may be due to the covariance structure $\hat{\Sigma}$ and its assumed distribution. The model may be refitted with various covariance structures and to obtain a model which is found to satisfactorily explain the observed data. If this is not achieved then joint normality of the error terms may be an unreasonable assumption indicating that the latent variable model may not be appropriate. It is possible to fit latent variable models which assume a different multivariate distribution for the error terms, however this is not considered here.

\clearpage

\section*{Appendix C}

\subsection*{Augmented binary method}

The augmented binary model is shown below. The baseline measures for $Y_{i1}$ and $Y_{i2}$ are included for comparison, as they are accounted for in the mean structure of the latent variable method. As one time point is modelled we can use a linear model for $Y_{i1}$ as shown in (\ref{augbincts}). Note that $Y_{i1}$ or $Y_{i2}$ may be chosen as the continuous measure to retain and should always be determined by which is the most informative.

\begin{equation}\label{augbincts}
Y_{i1}=\delta_{0}+\delta_{1} T{i}+\delta_{2} y_{i10}+\delta_{3} y_{i20}+\varepsilon_{i} \tag{C.1}
\end{equation}

where $\varepsilon_{i}|T_{i},y_{i10},y_{i20} \sim N(0,\sigma)$. In this case, the failure time binary indicator will contain information from the remaining three components. $F_{i}$ is set to equal 0 if $Y_{i2}\leq \theta_{2}, Y_{i3}$ is Grade B-E and $Y_{i4}=0$, otherwise the patient is labelled a non-responder in these components and $F_{i1}=1$. $F_{i}$ is modelled using the logistic regression model in (\ref{augbinbin}).

\begin{equation}\label{augbinbin}
logit(Pr(F_{i}=1\vert T_{i}, y_{i10},y_{i20})=\alpha_{F}+\beta_{F} T_{i}+\gamma_{F} y_{i10}+\psi_{F} y_{i20} \tag{C.2}
\end{equation}

Maximum likelihood estimates for the parameters are obtained from fitting models (\ref{augbincts}) and (\ref{augbinbin}). The probability of response 
is shown in (\ref{probrespaug}). 
\begin{equation} \label{probrespaug}
 P(Y_{i1}\leqslant\theta_{1},F_{i1}=0|T_{i},y_{i10},y_{i20}) \\ = \int_{-\infty}^{\theta_{1}}P(F_{i1}=0 | T_{i},y_{i10},y_{i20})f_{Y_{1}}(y_{i1};T_{i},y_{i10},y_{i20})dy_{i1} \tag{C.3}
\end{equation}\\
As in the latent variable method, (\ref{probrespaug}) is used to obtain probability of response estimates for each patient, assuming they were treated $\tilde{p_{i1}}$ and not treated $\tilde{p_{i1}}$, which are used to define an odds ratio, risk ratio or risk difference. 

\subsection*{Standard binary method}

The standard binary method is a logistic regression on the overall responder index, as shown in (\ref{standbin}). 
\begin{equation}\label{standbin}
logit(Pr(S_{i}=1\vert T_{i}, y_{i10})=\alpha+\beta T_{i}+\gamma y_{i10}+\psi y_{i20} \tag{C.4}
\end{equation}

The odds ratio and standard error estimates can be obtained directly. 

\section*{Appendix D}

\begin{table}[h!]
\caption{Performance measures and Monte Carlo standard errors used to assess the behaviour of the latent variable, augmented binary and binary methods in a simulation study for the systemic lupus erythematosus composite endpoint}\label{performancemeasures}
\centering
\begin{threeparttable}
\begin{tabular}{lll}
\hline
 \rule[-1ex]{0pt}{4ex}   Performance measure & Estimate & MCSE  \\
\hline \\
Bias & $\frac{1}{n_{sim}} \sum\limits_{j=1}^{n_{sim}}\hat{\theta}_{j} - \theta$ & $\sqrt{\frac{1}{n_{sim}(n_{sim}-1)} \sum\limits_{j=1}^{n_{sim}} (\hat{\theta}_{j} - \bar{\theta})^{2}}$ \\
Coverage & $\frac{1}{n_{sim}} \sum\limits_{j=1}^{n_{sim}}1(\hat{\theta}_{low,j}\leq \theta \leq \hat{\theta}_{upp,j})$& $\sqrt{\frac{\hat{cov.}(1-\hat{cov.})}{n_{sim}}}$\\
Bias-corrected coverage & $\frac{1}{n_{sim}} \sum\limits_{j=1}^{n_{sim}}1(\hat{\theta}_{low,j}\leq \bar{\theta} \leq \hat{\theta}_{upp,j})$& $\sqrt{\frac{\hat{BEcov.}(1-\hat{BEcov.})}{n_{sim}}}$\\
Power & $\frac{1}{n_{sim}} \sum\limits_{j=1}^{n_{sim}}1(p_{j}<\alpha)$ & $\sqrt{\frac{\hat{Power}(1-\hat{Power})}{n_{sim}}}$ \\
MSE &$\sum\limits_{j=1}^{n_{sim}} (\hat{\theta}_{j} - \theta)^{2}$ & $\sqrt{\frac{\sum\limits_{j=1}^{n_{sim}}[(\hat{\theta}_{j} - \theta)^{2}-\hat{MSE}]^{2}}{n_{sim}(n_{sim}-1)}}$ \\
Empirical SE &  $\sqrt{\frac{1}{n_{sim}-1} \sum\limits_{j=1}^{n_{sim}} (\hat{\theta}_{i} - \bar{\theta})^{2}}$ & $\frac{\hat{EmpSE}}{\sqrt{2(n_{sim}-1)}}$ \\
Model SE &$\sqrt{\frac{1}{n_{sim}-1} \sum\limits_{j=1}^{n_{sim}}\hat{Var}(\hat{\theta}_{j})}$ & $\sqrt{\frac{\hat{Var}[\hat{Var}(\hat{\theta})]}{4n_{sim}\hat{ModSE}^{2}}} \dagger$\\
Relative precision A vs. B & $ \frac{\hat{Var}(\hat{\theta}_{j})_{B}}{\hat{Var}(\hat{\theta}_{j})_{A}}$ & - \\
& & \\
\hline\\
\end{tabular}
\begin{tablenotes}
\item[]$\hat{\theta}_{j}:$ estimated log-odds treatment effect in simulated data j
\item[]$\bar{\theta}$: mean log-odds treatment effect over $n_{sim}$ datasets
\item[] $\hat{\theta}_{low,j}, \hat{\theta}_{upp,j}$ lower and upper limit of confidence interval for iteration j
\item[$\dagger$] $\hat{Var}[\hat{Var}(\hat{\theta})]=\frac{1}{n_{sim}-1}\sum_{j=1}^{n_{sim}}\lbrace \hat{Var}(\hat{\theta}_{i})-\frac{1}{n_{sim}}\sum_{j=1}^{n_{sim}}\hat{Var}(\hat{\theta}_{j})\rbrace ^{2}$
\end{tablenotes}
\end{threeparttable}
\end{table}

\section*{Appendix E}

\begin{table}[h!]
\caption{Parameter values for the simulated scenarios which investigate the effect of varying responder threshold $\theta_{1}$, changing the components driving response and differing treatment effects on the performance of the latent variable, augmented binary and standard binary methods for the systemic lupus erythematosus composite endpoint}\label{simvary}
\begin{center}
\begin{tabular}{ccc}
\hline
\rule[-1ex]{0pt}{4ex}  Scenario& Parameters& Investigates\\
\hline
&&\\
$\theta_{1}=-2$ & $\theta_{1}=-2$ & 100\% of patients respond in $Y_{1}$\\
$\theta_{1}=-3$ & $\theta_{1}=-3$  & 96\% of patients respond in $Y_{1}$ \\
$\theta_{1}=-4$ & $\theta_{1}=-4$ & 82\% of patients respond in $Y_{1}$ \\
$\theta_{1}=-5$ &$\theta_{1}=-5$  & 52\% of patients respond in $Y_{1}$\\
$\theta_{1}=-6$ & $\theta_{1}=-6$ & 20\% patients respond in $Y_{1}$\\
$Y_{1}, Y_{4}$ & $\theta_{1}=-5, \theta_{2}=2, \theta_{3}=2$ & \makecell{Continuous and binary variable \\ driving response }\\
$Y_{4}$ & $\theta_{1}=-2, \theta_{2}=2, \theta_{3}=2$ & \makecell{Binary variable driving \\response}\\
$Y_{1}, Y_{2}, Y_{3}$ & $\theta_{4}=2$ & \makecell{Two continuous and ordinal \\drive response}\\
Treat case 1 & \makecell{$\alpha_{0}=-4.9, \alpha_{1}=-0.09,\beta_{0}=-1.2,$\\$\beta_{1}= -0.11, \gamma_{1}=-0.145, \psi_{0}=-0.2,$\\ $\psi_{1}=-0.07$} &Odds ratio = 1.217\\
Treat case 2 & \makecell{$ \alpha_{0}=-4.9, \alpha_{1}=-0.20,\beta_{0}=-1.2, $\\$\beta_{1}= -0.25, \gamma_{1}=-0.2, \psi_{0}=-0.2, $\\$\psi_{1}=-0.12$  }& Odds ratio = 1.426\\
Treat case 3 & \makecell{$\alpha_{0}=-4.9, \alpha_{1}=-0.30, \beta_{0}=-1.2, $\\$ \beta_{1}= -0.50, \gamma_{1}=-0.3, \psi_{0}=-0.2, $\\$\psi_{1}=-0.22$ }& Odds ratio = 1.794 \\
Treat case 4 & \makecell{$\alpha_{0}=-4.9, \alpha_{1}=-0.32, \beta_{0}=-1.2, $\\$\beta_{1}= -0.65, \gamma_{1}=-0.39, \psi_{0}=-0.2, $\\$\psi_{1}=-0.27$} & Odds ratio = 2.007 \\
Treat case 5 & \makecell{$\alpha_{0}=-4.9, \alpha_{1}=-0.33, \beta_{0}=-1.2, $ \\$\beta_{1}= -0.72,\gamma_{1}=-0.45, \psi_{0}=-0.2, $\\$\psi_{1}=-0.33$} & Odds ratio = 2.198\\
&&\\
\hline\\
\end{tabular}
\end{center}
\end{table}

\section*{Appendix F}

\subsection*{Multivariate skew-normal distribution}
To test the robustness of the latent variable method to deviations from joint normality of the components, we can generate the data so that the components are drawn from a multivariate skew-normal. The multivariate skew-normal is an extension of the univariate skew-normal distribution introduced by \cite{multskew}. They define it as follows. A random vector \textbf{Y}=$(Y_{1},...,_{k})^{T}$ has k-variate skew-normal distribution, if its density function is \begin{equation}\label{multskew}
f_{k}(\mathbf{y} )=2\phi_{k}(\mathbf{y;\Omega})\Phi(\boldsymbol{\alpha} ^{T}\mathbf{y}),   \mathbf{y}\in \mathbf{R}^{k}\tag{F.1} \end{equation} where $\phi_{k}(\mathbf{y;\Omega})$ is the probability density function of the k-variate normal distribution with standardised marginals and correlation matrix $\mathbf{\Omega}$. The shape parameter $\boldsymbol{\alpha}$ determines the skewness, where $\boldsymbol{\alpha}=\mathbf{0}$ reduces the density in (\ref{multskew}) to the N($\mathbf{0},\boldsymbol{\Omega}$) density.\\
Scenarios of interest are shown in Table \ref{skew}. The first scenario considers when all four components are skewed.  Scenarios 2-3 consider different magnitudes of skew in the latent continuous components only. This tests the robustness of the method to the assumption that the observed discrete variables manifest from continuous variables. Scenario 4 is the null case for scenario 3. 

\begin{table}[h!]
\caption{Simulation scenarios considered to investigate deviations from joint normality for the components of the systemic lupus erythematosus composite endpoint based on the multivariate skew-normal distribution where $\boldsymbol{\alpha}$ determines the magnitude of the skew in each component}\label{skew}
\centering
\begin{tabular}{lll}
\hline
\rule[-1ex]{0pt}{4ex} Scenario &  $\boldsymbol{\alpha}$ & Purpose \\
\hline
&&\\
\centering skew1 & \centering (0.1, 0.1, 0.1, 0.1) & Skew in all four components \\
\centering skew2 & \centering (0, 0, 0.1, 0.1) & Skew in discrete components only	\\
\centering skew3 & \centering  (0, 0, 0.05, 0.05) & Smaller skew in discrete components only \\
\centering skew4 & \centering  (0, 0, 0.05, 0.05) & Smaller skew in discrete components only in the null case\\
&&\\
\hline\\
\end{tabular}
\end{table}

\subsection*{Results}

The bias, coverage, bias-corrected coverage and power are shown in Table \ref{skewop} for all four scenarios. In scenarios 1-3, the non-normality introduces bias which results under-coverage. The bias-corrected coverage is close to nominal for all scenarios however the coverage of the latent variable method is nominal in the null case. This is consistent with our findings when the joint normality assumption is satisfied in that bias is introduced in the estimation of the treatment arm, however the magnitude of this bias is much smaller when the assumptions are satisfied. The augmented binary and standard binary methods behave similarly to when the joint normality assumptions are satisfied, which is expected given that the assumptions of those models are violated in both contexts. The latent variable method still offers large power gains over the other methods. 

\begin{table}[h!]
\caption{Operating characteristics of the latent variable, augmented binary and binary methods when the components of the systemic lupus erythematosus endpoint are drawn from a multivariate skew-normal, N=300 and $n_{sim}=1000$}
\label{skewop}
\centering
\begin{tabular}{ccccc}
\hline
\rule[-1ex]{0pt}{4ex}  Performance measure  & Scenario &\multicolumn{3}{c}{Method} \\
 \cmidrule(r){3-5}
\rule[-1ex]{0pt}{4ex}   &  &  Latent Variable & Augmented Binary & Binary\\  \hline
 \hspace{0.6cm}  & & &  & \\
\hspace{0.6cm} Bias & skew1 & -0.173 (0.012) & 0.041 (0.252) & -0.015 (0.258)\\
\hspace{0.6cm} & skew2 & -0.103 (0.008) & 0.036 (0.251) & -0.020 (0.255) \\
\hspace{0.6cm} & skew3 & -0.068 (0.008) & 0.038 (0.244) & -0.016 (0.245) \\
\hspace{0.6cm} & skew4 &  -0.033 (0.008) & 0.007 (0.254) & 0.001 (0.255) \\
\hspace{0.6cm}  & & &  & \\
\hspace{0.6cm} Coverage& skew1 &  0.556 (0.018) & 0.933 (0.009) & 0.939 (0.009) \\
\hspace{0.6cm} & skew2 &0.811 (0.013) & 0.928 (0.008) & 0.941 (0.008) \\
\hspace{0.6cm} & skew3 & 0.884 (0.010) & 0.934 (0.008) & 0.950 (0.007) \\
\hspace{0.6cm} & skew4 & 0.933 (0.009) & 0.923 (0.009) & 0.950 (0.008) \\
\hspace{0.6cm}  & & &  & \\
\hspace{0.6cm} Bias-corrected & skew1 & 0.962 (0.007) & 0.929 (0.009) & 0.943 (0.008) \\
\hspace{0.6cm} coverage & skew2 & 0.936 (0.008) & 0.930 (0.008) & 0.943 (0.007) \\
\hspace{0.6cm} & skew3 &  0.940 (0.008) & 0.929 (0.008) & 0.954 (0.007) \\
\hspace{0.6cm} & skew4 & 0.948 (0.008) & 0.926 (0.009) & 0.950 (0.008) \\
\hspace{0.6cm}  & & &  & \\
\hspace{0.6cm} Power & skew1 & 0.897 (0.011) & 0.646 (0.017) & 0.487 (0.018) \\
\hspace{0.6cm} & skew2 &  0.959 (0.006) & 0.637 (0.015) & 0.471 (0.016) \\
\hspace{0.6cm} & skew3 &0.982 (0.004) & 0.641 (0.015) & 0.495 (0.016) \\
\hspace{0.6cm} & skew4 &- &- &- \\
\hspace{0.6cm}  & & &  & \\
\hline
\end{tabular}
\end{table}

Table \ref{SEs} shows the MSE, empirical SE and model SE of the three methods. The latent variable method performs best consistently across these performance measures. The augmented binary and standard binary methods have an MSE across all scenarios of approximately 0.06 whilst the MSE of the latent variable method is between 0.01 and 0.04. This indicates that the large reduction in variance is useful despite the introduction of bias. We acknowledge however that this may not hold across all sample sizes (\cite{Simstudy}).

\begin{table}[h!]
\caption{Operating characteristics (Monte Carlo standard errors in parentheses) of the latent variable, augmented binary and binary methods when the components of the systemic lupus erythematosus endpoint are drawn from a multivariate skew-normal, N=300 and $n_{sim}=1000$}
\label{SEs}
\centering
\begin{tabular}{ccccc}
\toprule
Performance measure  & Scenario &\multicolumn{3}{c}{Method} \\
 \cmidrule(r){3-5}
 &  &  Latent Variable & Augmented Binary & Binary\\  \hline
 & & & & \\
 \hspace{0.6cm}  MSE & skew1 &  0.039 (0.001) & 0.063 (0.003) & 0.066 (0.003) \\
\hspace{0.6cm} &skew2 & 0.021 (0.001) & 0.063 (0.003) & 0.065 (0.003) \\
\hspace{0.6cm} &skew3 & 0.014 (0.001) & 0.060 (0.003) & 0.060 (0.003) \\
\hspace{0.6cm} &skew4 & 0.010 (0.001) & 0.064 (0.004) & 0.065 (0.003)\\
\hspace{0.6cm}   & &  & \\
\hspace{0.6cm} EmpSE & skew1 &0.097 (0.003) & 0.248 (0.006) & 0.257 (0.007) \\
\hspace{0.6cm} & skew2 & 0.102 (0.002) & 0.249 (0.006) & 0.254 (0.006) \\
\hspace{0.6cm} &skew3 & 0.099 (0.002) & 0.241 (0.005) & 0.245 (0.006) \\
\hspace{0.6cm} &skew4 & 0.094 (0.002) & 0.254 (0.006) & 0.255 (0.006) \\
\hspace{0.6cm}  & & &  & \\
\hspace{0.6cm} ModSE & skew1 & 0.010 (0.006) & 0.052 (0.001) & 0.064 (0.001)  \\
\hspace{0.6cm} & skew2 &0.010 (0.003) & 0.050 (0.001) & 0.060 (0.001) \\
\hspace{0.6cm} & skew3 & 0.010 (0.015) & 0.048 (0.001) & 0.059 (0.001) \\
\hspace{0.6cm} &skew4 & 0.009 (0.004) & 0.051 (0.001) & 0.063 (0.001) \\
\hspace{0.6cm}  & & &  & \\
\hline
\end{tabular}\\
\end{table}

Table \ref{skewprob} shows the probability of response in each arm for each of the methods. The findings are consistent with when the assumptions are satisfied. Namely, the latent variable method estimates the probability of response in the control arm well however underestimates the probability of response in the treatment arm. The magnitude of this underestimation is unaffected by the degree of skew or whether the skew is present in the observed continuous components. 

\begin{table}[h!]
\caption{Estimated probability of response in the treatment and placebo arms from the latent variable model (Lat Var), augmented binary method (Aug Bin) and standard binary method (Bin) when the components of the systemic lupus erythematosus endpoint are drawn from a multivariate skew-normal, N=300 and $n_{sim}=1000$}
\label{skewprob}
\centering
\begin{tabular}{ccccccccc}
\toprule
&\multicolumn{4}{c}{$Pr(resp\mid T=0)$} & \multicolumn{4}{c}{$Pr(resp\mid T=1)$} \\
 \cmidrule(r){2-5}\cmidrule(r){6-9}
Scenario &True & Lat Var &  Aug Bin &Bin & True& Lat Var &  Aug Bin &  Bin \\  \hline
& & & & & & & \\
\hspace{0.6cm} skew1 &  0.259 & 0.263 & 0.221 & 0.258 & 0.365 & 0.330 & 0.326 & 0.359\\
\hspace{0.6cm} skew2 & 0.290 & 0.287 & 0.253 & 0.290 & 0.398 & 0.370 & 0.361 & 0.392 \\
\hspace{0.6cm} skew3 &  0.309 & 0.302 & 0.271 & 0.308 & 0.418 & 0.394 & 0.382 & 0.413\\
 \hspace{0.6cm} skew4 &  0.309 & 0.299 & 0.269 & 0.307 & 0.309 & 0.292 & 0.270 & 0.307\\
\hspace{0.6cm} & &  & &  &  &  &  &  \\
\hline
\end{tabular}\\
\end{table}

The odds ratio treatment effect estimate from each method is shown in Table \ref{treateffect}. The latent variable method is biased towards the null, the augmented binary method is biased away from the null. The binary method slightly underestimates the treatment effect in this setting however all are close to true for the null case.

\begin{table}[h!]
\caption{Estimated odds ratio treatment effect from the latent variable model (Lat Var), augmented binary method (Aug Bin) and standard binary method (Bin) when the components of the systemic lupus erythematosus endpoint are drawn from a multivariate skew-normal, N=300 and $n_{sim}=1000$}
\label{treateffect}
\centering
\begin{tabular}{ccccc}
\toprule
&\multicolumn{4}{c}{Treatment effect} \\
 \cmidrule(r){2-5}
Scenario &True & Lat Var &  Aug Bin &Bin \\  \hline
& & & &  \\
\hspace{0.6cm} skew1 &1.640 & 1.379 (1.140, 1.668) & 1.708 (1.093, 2.668) & 1.616 (0.985, 2.651) \\
\hspace{0.6cm} skew2 &  1.617 & 1.459 (1.203, 1.770) & 1.676 (1.083, 2.594) & 1.586 (0.980, 2.565) \\
\hspace{0.6cm} skew3 & 1.611 & 1.505 (1.243, 1.822) & 1.674 (1.089, 2.572) & 1.585 (0.987, 2.548) \\
\hspace{0.6cm} skew4 &  1.000 & 0.967 (0.807, 1.160) & 1.007 (0.647, 1.566) & 1.001 (0.613, 1.634)\\
\hspace{0.6cm} & &  & &   \\
\hline
\end{tabular}\\
\end{table}

The median relative precision of the methods are shown in Table \ref{skewprec}, with the 10th centile and 90th centile values. These are consistent with our previous findings indicating that the violation of joint normality only affects the bias and not the variance. 

\begin{table}[h!]
\caption{Estimated relative precision from the latent variable model (Lat Var), augmented binary method (Aug Bin) and standard binary method (Bin) when the components of the systemic lupus erythematosus endpoint are drawn from a multivariate skew-normal, N=300 and $n_{sim}=1000$}
\label{skewprec}
\centering
\begin{tabular}{cccc}
\toprule
&\multicolumn{3}{c}{Treatment effect} \\
 \cmidrule(r){2-4}
Scenario &Lat Var vs Bin & Lat Var vs Aug Bin & Aug Bin vs Bin \\  \hline
& & & \\
\hspace{0.6cm} skew1 & 6.903 [5.336, 8.972] & 5.579 [4.376, 7.313] & 1.231 [1.189, 1.275]  \\
\hspace{0.6cm} skew2 &  6.263 [5.013, 7.917] & 5.177 [4.096, 6.518] & 1.213 [1.178, 1.252] \\
\hspace{0.6cm} skew3 & 6.326 [5.016, 7.995] & 5.192 [4.098, 6.548] & 1.219 [1.184, 1.257]\\
\hspace{0.6cm} skew4 &7.384 [5.729, 9.343] & 5.985 [4.655, 7.629] & 1.231 [1.192, 1.273]\\
\hspace{0.6cm} & &  &    \\
\hline
\end{tabular}\\
\end{table}

\clearpage
\section*{Appendix G}

\begin{figure}[h!]
\centering \includegraphics[scale=1]{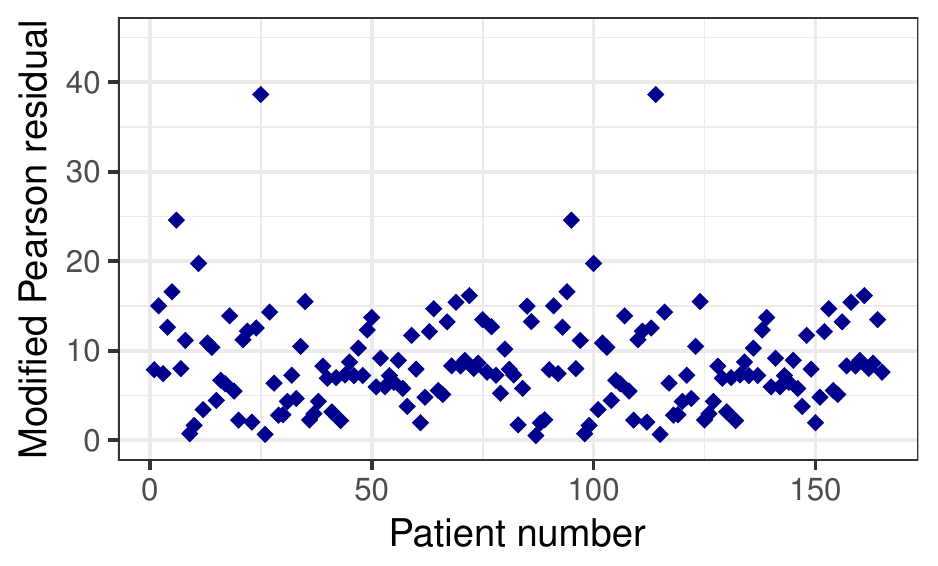}\caption{Plot of the modified Pearson residuals from the latent variable model for each patient in the MUSE trial. The residuals highlight that two patients observations are poorly explained by the model but that the model is a good fit for the remaining patients.}\label{modifiedpearson}
\end{figure}  

\begin{figure}[h!]
\centering \includegraphics[scale=1]{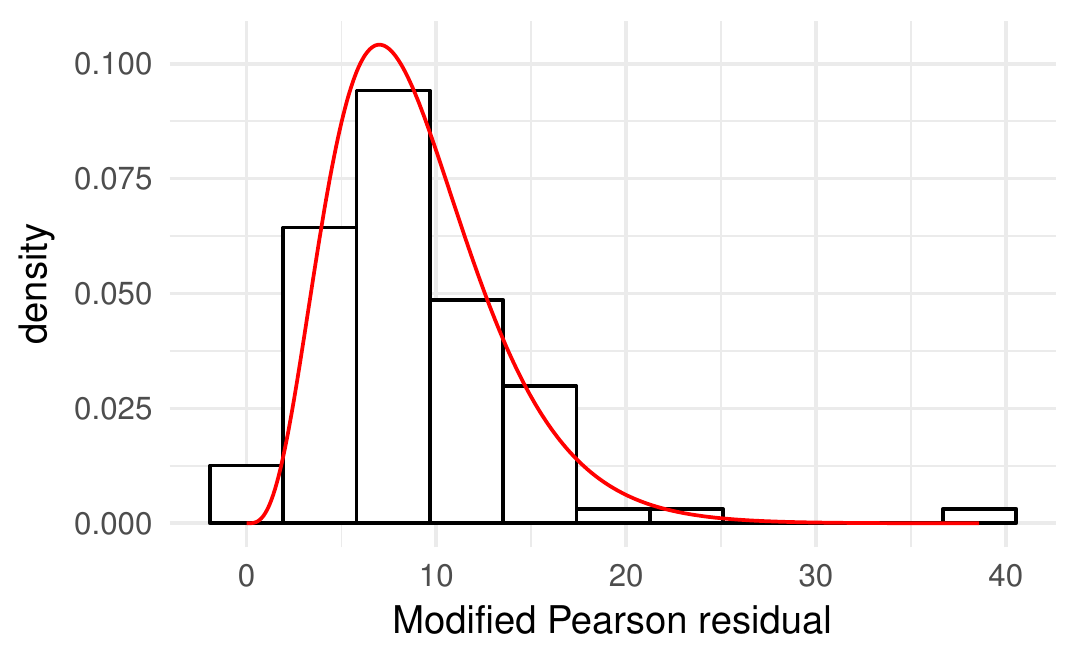}\caption{Histogram of the modified Pearson residuals from the latent variable model in the MUSE trial dataset with the corresponding $\chi^{2}$ density. The modified Pearson residuals should follow the distribution of the $\chi^{2}$ density shown if the model fits well.}\label{chisq}
\end{figure}

\end{document}